\documentclass[12pt]{article}

\pdfoutput=1

\usepackage{amsmath,amssymb,enumerate}
\usepackage{graphicx,color}
\usepackage{wasysym}
\usepackage{lipsum}
\usepackage{float}
\usepackage{cancel}
\usepackage{empheq}
\usepackage{caption}
\usepackage{subcaption}
\usepackage{geometry}
\usepackage{titling}
\usepackage{textpos}

\newcommand{\om}[0]{\omega}

\newcommand{\amp}{&\!\!}
\newcommand{\beq}{\begin{equation}}
\newcommand{\eeq}{\end{equation}}
\newcommand{\bea}{\begin{eqnarray}}
\newcommand{\eea}{\end{eqnarray}}
\newcommand{\mpl}{M_{Pl}}
\newcommand{\aaa}{\tilde{\alpha}}

\begin{document}
\title{\bf Oscillatory Attractors: \\ A New Cosmological Phase}

\author{
\Large{Jasdeep S.~Bains$^{1,2}$, Mark P.~Hertzberg$^{3,4}$, Frank Wilczek$^4$} \\
~\\
{\em $^1$Center for the Fundamental Laws of Nature},\\
{\em Harvard University, Cambridge, MA 02138, USA}\\
\\
{\em $^2$Perimeter Institute for Theoretical Physics},\\
{\em Waterloo, ON N2L 2Y5, Canada}\\
\\
{\em $^2$Institute of Cosmology, Department of Physics and Astronomy},\\
{\em Tufts University, Medford, MA 02155, USA}\\
\\
{\em $^3$Center for Theoretical Physics, Department of Physics},\\
{\em MIT, Cambridge, MA 02139, USA}
}

\date{\today}

\maketitle 

\begin{textblock*}{5cm}(15cm,-12cm)
\fbox{\footnotesize MIT-CTP-4745}
\end{textblock*}

\vspace{-0.8cm}

\begin{abstract}
In expanding FRW spacetimes, it is usually the case that homogeneous scalar fields redshift and their amplitudes approach limiting values: Hubble friction usually ensures that the field relaxes to its minimum energy configuration, which is usually a static configuration. Here we discover a class of relativistic scalar field models in which the attractor behavior is the field oscillating indefinitely, with finite amplitude, in an expanding FRW spacetime, despite the presence of Hubble friction. This is an example of spontaneous breaking of time translation symmetry. We find that the effective equation of state of the field has average value $\langle w\rangle=-1$, implying that the field itself could drive an inflationary or dark energy dominated phase. This behavior is reminiscent of ghost condensate models, but in the new models, unlike in the ghost condensate models, the energy-momentum tensor is time dependent, so that these new models embody a more definitive breaking of time translation symmetry. We explore (quantum) fluctuations around the homogeneous background solution, and find that low $k$-modes can be stable, while high $k$-modes are typically unstable. We discuss possible interpretations and implications of that instability.
\let\thefootnote\relax\footnotetext{Email: {\tt bains@physics.harvard.edu/@pitp.ca, mark.hertzberg@tufts.edu, wilczek@mit.edu}}
\end{abstract}

\newpage

\tableofcontents

\newpage

\section{Introduction}

The phenomenon of spontaneous symmetry breaking is an essential idea in many areas of physics, including magnetisation, crystallisation, relativistic field theory, etc. In the case of crystallisation, the atoms arrange themselves in the ground state into a definite periodic lattice defined throughout space, which spontaneously breaks spatial translation symmetry. An intriguing question is whether analogous behavior can occur in the time direction, i.e., whether or not systems exist whose ground states spontaneously break time translation symmetry. Furthermore, if this breaking were to be periodic in time, then it would be a kind of ``time crystal". This idea was introduced and explored in several recent papers \cite{Shapere:2012nq,Wilczek:2012jt}, with related work in Refs.~\cite{Shapere:2012yf,Wilczek:2013,Shapere:2012wn,Sacha:2015tga,Chi:2013sea,Curtright:2013nya}.

This idea is difficult to realize for at least three reasons: (i) in classical mechanics, one has the Hamilton equations $\dot p=-\partial H/\partial x,\, \dot x=\partial H/\partial p$. Since the ground state is an extremum of the Hamiltonian, we expect $\partial H/\partial x=\partial H/\partial p=0$ and therefore $\dot x=\dot p=0$ in the ground state, and hence no breaking of the time translation symmetry. To avoid this, one needs to introduce Hamiltonians that are not smooth functions of their arguments.  One may have extrema at cusps, where the partial derivatives of $H$ do not vanish, allowing the possibility of motion.  However, this leads to a second difficulty: (ii) since the Hamiltonians are not smooth at their minima, the ground state behavior can be pathological or ill defined.  If we pass to the quantum theory, the difficulty re-emerges as follows: (iii) the Heisenberg equation of motion for some observable $\mathcal{O}$ states $\dot{\mathcal{O}}=i[\hat{H},\mathcal{O}]/\hbar$, and since one expects the ground state to be an eigenstate of $\hat{H}$, then the expectation value of the commutator gives $\langle\dot{\mathcal{O}}\rangle=0$, again implying no motion.  All in all, realizing spontaneous time translation symmetry breaking is challenging.  However we will now sketch a way to finesse those difficulties, which will be fully documented in what follows.

In this work, we will focus on a relativistic scalar field $\phi$ and discuss the possibility of spontaneously breaking time translation symmetry in this context. To avoid difficulty (i) we will need a theory with a cusp in its Hamiltonian. This can be achieved by starting with a smooth Lagrangian, with a wrong sign quadratic kinetic term, supplemented by higher order kinetic terms. Such smooth, albeit unconventional, Lagrangians generally give rise to non-smooth Hamiltonians, as we will discuss.  To avoid difficulty (ii), in this paper we will consider our scalar field in an expanding FRW universe. In this context Hubble expansion can generate an effective friction, that might be expected to drive a system to its ground state.   We will find, however, that in many circumstances our field in fact evolves to a different state, which does not approach the Hamiltonian's cusp.  In this attractor configuration we find that the field oscillates, in the homogeneous approximation, indefinitely and periodically.  We will therefore have true breaking of the time translation symmetry, and an associated ``time crystal" behavior. The attractor configuration is a kind of vacuum in cosmology, so we feel justified in referring to this phenomenon as spontaneous symmetry breaking.  We note that although the attractor behavior avoids the cusp, this does not necessarily remove the need for the cusp in the Hamiltonian; the Hamiltonian is generally multi-valued in this context, and the attractor solution occurs at a point in phase space where there is a crossing - this usually leads to a cusp at another point in phase space, though there may be special cases where this is avoided altogether.
Finally, difficulty (iii) is not manifested in any direct way, since the field theory is treated classically, as is appropriate to large occupation numbers.   We will later consider quantum fluctuations, and find that this introduces additional challenges, which we will discuss.

In this paper, we analyze a class of models that exhibit this oscillatory behavior.  Specifically, we consider a relativistic scalar field $\varphi$ with a $\varphi\to-\varphi$ symmetry, which is described by an effective field theory.
The most general form of the Lagrangian, which only propagates a single degree of freedom, is built out of first derivatives of $\varphi$, which we write as follows ($+---$ signature)
\beq
\mathcal{L}=\mathcal{L}(\varphi,\partial\varphi)=\sum_{n,m} c_{n,m}\varphi^{2n}(\partial\varphi)^{2m}.
\eeq
For the leading set of operators, we make the following choice of signs: $c_{0,1}<0$, $c_{0,2}>0$, $c_{1,0}>0$, $c_{2,0}<0$, $c_{1,1}>0$.
These choices of signs lead to a kind of symmetry breaking Lagrangian in both the $\varphi$ direction and the $\partial\varphi$ direction in field space.

It is well known that the choice $c_{1,0}>0$, $c_{2,0}<0$ leads to a double well type of potential. On the other hand, the ghost condensate models arise from $c_{n,0}=0$, and having only kinetic terms with $c_{0,1}<0$, $c_{0,2}>0$. 
This leads to a so-called ghost condensate, whereby the field organizes into a configuration with $\dot\varphi$  a non-zero constant \cite{ArkaniHamed:2003uy,ArkaniHamed:2003uz}. This solution breaks the most straightforward time translation invariance, since $\varphi\propto t$, but the energy-momentum tensor is independent of time (since it is only a function of derivatives of $\varphi$).  Indeed, a combined field shift-time translation operation remains a symmetry of the solution. Instead, in this paper we allow not only for a negative kinetic term but also for a double well potential.  This supports oscillatory behavior for $\varphi$ in the ground state and, more importantly for our present purposes, in the attractor state in an expanding universe.  The corresponding energy-momentum tensor is truly time dependent.

We will explore homogeneous but time-dependent solutions of this class of Lagrangians systematically, over a broad parameter space.  We show that in general four behaviors are possible, where the attractor solutions are (a) oscillatory, (b) constant, (c) vanishing, and (d) singular. The first class, (a) oscillatory, offers a dramatic example of spontaneous breaking of time translation symmetry and leads to potentially interesting cosmologies. We show that the effective equation of state for the oscillatory field is $\langle w\rangle=-1$, so that it can drive a phase of inflation or be manifested as dark energy.  We also find that since the background is oscillatory, fluctuations around it can exhibit parametric resonance, depending on parameters and wavenumber. We find a regime in which the long wavelength modes are stable, and show that the short wavelength modes typically grow exponentially in time. 

Our paper is organized as follows: 
In Section \ref{model} we describe the model in more detail and give an overview of the behavior of solutions.
In Section \ref{ab1} we provide  analytical results for a special choice of parameters.
In Section \ref{closeto11} we provide more general results for quasi-special parameters and then completely arbitrary parameters.
In Section \ref{fluct} we include fluctuations around the homogeneous vacuum solution, both numerically and analytically.
In Section \ref{Cosmology} we discuss possible cosmological implications.
In Section \ref{Conclude} we summarize and draw conclusions.
Finally, Appendix \ref{HigherOrder} contains supplementary calculations.

\section{Model}\label{model}

For our purposes it will suffice to confine attention to a few operators that are relevant to describe the vacuum, namely $c_{0,0}, c_{0,1}, c_{1,0}, c_{2,0}, c_{1,1}, c_{0,2}$.  We can view these as the leading terms in an effective theory.  Later we will describe the plausible regime of validity of this effective theory.

\subsection{General}

We begin with six parameters, but by re-scaling the unit of field strength $\varphi\to\phi$ and the unit of length, and taking into account that an overall multiplicative factor does not affect the classical equations of motion, we can capture the possible behaviors using just three parameters $a,b,\Lambda$, using a Lagrangian of the form 
\beq\label{lagrangian}
\mathcal{L}=
-\Lambda-{1\over12a}
+{1\over2}\phi^2
-\frac{3 \,a}{4}\phi^4
- \frac{1}{2}(\partial\phi)^2
+ \frac{1}{4} (\partial\phi)^4
+\frac{3\,b}{2}\phi^2 (\partial\phi)^2.
\eeq
We have organized the constant terms such that the minimum of the potential is $\Lambda$.

From the Lagrangian we derive the canonical momentum field
\beq
\pi_\phi=\frac{\partial\mathcal{L}}{\partial\dot{\phi}}=-\dot\phi+\dot\phi(\partial\phi)^2+3\,b\,\phi^2\dot{\phi}.
\eeq
To invert this equation, solving for $\dot \phi$, we must solve a cubic equation.  Doing that generally involves multiple solutions.   
The Hamiltonian density, written in terms of $\phi$ and $\dot\phi$ is
\begin{eqnarray}
\mathcal{H}
\amp=\amp\pi_\phi\,\dot\phi-\mathcal{L}\label{ham1}\\
\amp=\amp
\Lambda
+{1\over12a}
-\frac{1}{2}\phi^2
+\frac{3}{4}a\phi^4
-\frac{1}{2}\left(\dot{\phi}^2
+(\nabla\phi)^2\right)\nonumber\\
\amp\amp\!\!+\frac{3}{4}\dot{\phi}^4
+\frac{1}{2}\dot{\phi}^2(\nabla\phi)^2
-\frac{1}{4}(\nabla\phi)^4
+\frac{3}{2}b\,\phi^2\left(\dot{\phi}^2
+(\nabla\phi)^2\right).\label{ham2}
\end{eqnarray}
When we rewrite this in terms of the canonical variables $\phi$ and $\pi_\phi$, we find {\em cusps} in the Hamiltonian density due to the multivalued nature of $\dot\phi=\dot\phi(\pi_\phi,\phi)$.  This avoids difficulty (i). 

In this work we take $a>0$ and $b>0$, which makes many of the contributions to $\mathcal{H}$ positive, and avoids any catastrophic instabilities for homogeneous fields.  There remains, however, a source of unboundedness from below, at large wave numbers, stemming from the $-1/4(\nabla\phi)^4$ term.  We can imagine that the Lagrangian is extended by higher order terms, such as $\sim(\partial\phi)^6$, etc, to fix that instability \cite{Shapere:2012nq}.  We defer further discussion of that complication pending our treatment of deviations from spatial homogeneity in  Section \ref{fluct}.

\subsection{Background Evolution}

In the first several sections of this paper we shall consider homogeneous evolution, where $\phi=\phi(t)$ is a function of time only.
We will consider the behavior of such fields an expanding FRW flat universe. The corresponding action is then
\beq
S=\int d^4x\,R^3\,\mathcal{L}(\phi,\dot\phi),
\eeq
where $R=R(t)$ is the scale factor of the expanding universe.  In our detailed work we will assume a de Sitter background, characterised by a constant Hubble parameter $H=\dot R/R$.   As we will see, $\phi$ itself can provide an cosmological constant or dark energy term, contributing to de Sitter evolution self-consistently.   

The homogenous energy density and pressure are given by
\begin{eqnarray}
\label{energy}
\rho&=&
+\Lambda+\frac{1}{12a}
-\frac{1}{2}\phi^2
+\frac{3}{4}a\phi^4
-\frac{1}{2}\dot{\phi}^2
+\frac{3}{4}\dot{\phi}^4
+\frac{3}{2}b\,\phi^2\dot{\phi}^2,\\
p&=&
-\Lambda-\frac{1}{12a}
+\frac{1}{2}\phi^2
-\frac{3}{4}a\phi^4
-\frac{1}{2}\dot{\phi}^2
+\frac{1}{4}\dot{\phi}^4
+\frac{3}{2}b\,\phi^2\dot{\phi}^2.
\label{pressure}
\end{eqnarray}

By varying the action with respect to $\phi$ we obtain the equation of motion 
\beq\label{eomfrac}
(-1+3\dot{\phi}^2+3b\phi^2 ) \, \ddot{\phi}= \phi\left(1-3a\phi^2-3b\dot{\phi}^2\right)+3H\dot{\phi}\left(1-\dot{\phi}^2-3b\phi^2\right)
\eeq
including Hubble friction.  Note that the coefficient of $\ddot\phi$, $-1+3\dot{\phi}^2+3b\phi^2 $, can vanish, potentially leading to singular behavior (if the right hand side does not vanish).

\subsection{Overview of Solutions}\label{overview}

In subsequent sections we investigate in some detail how the nature of the solution landscape depends upon the initial conditions of the field, the Hubble parameter and the Lagrangian parameters $a$ and $b$. In this section we survey the qualitative forms our solutions settle into.

\begin{figure}[H]
\centering
\includegraphics[scale=0.52]{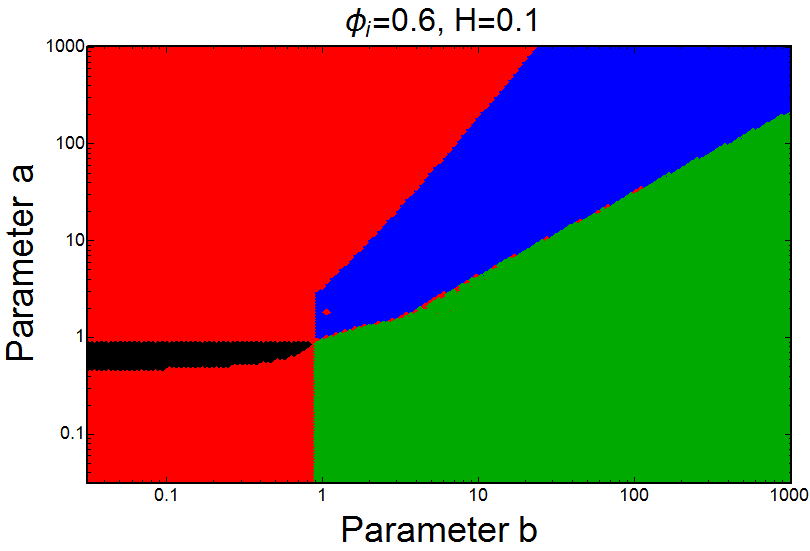}\\
\includegraphics[scale=0.25]{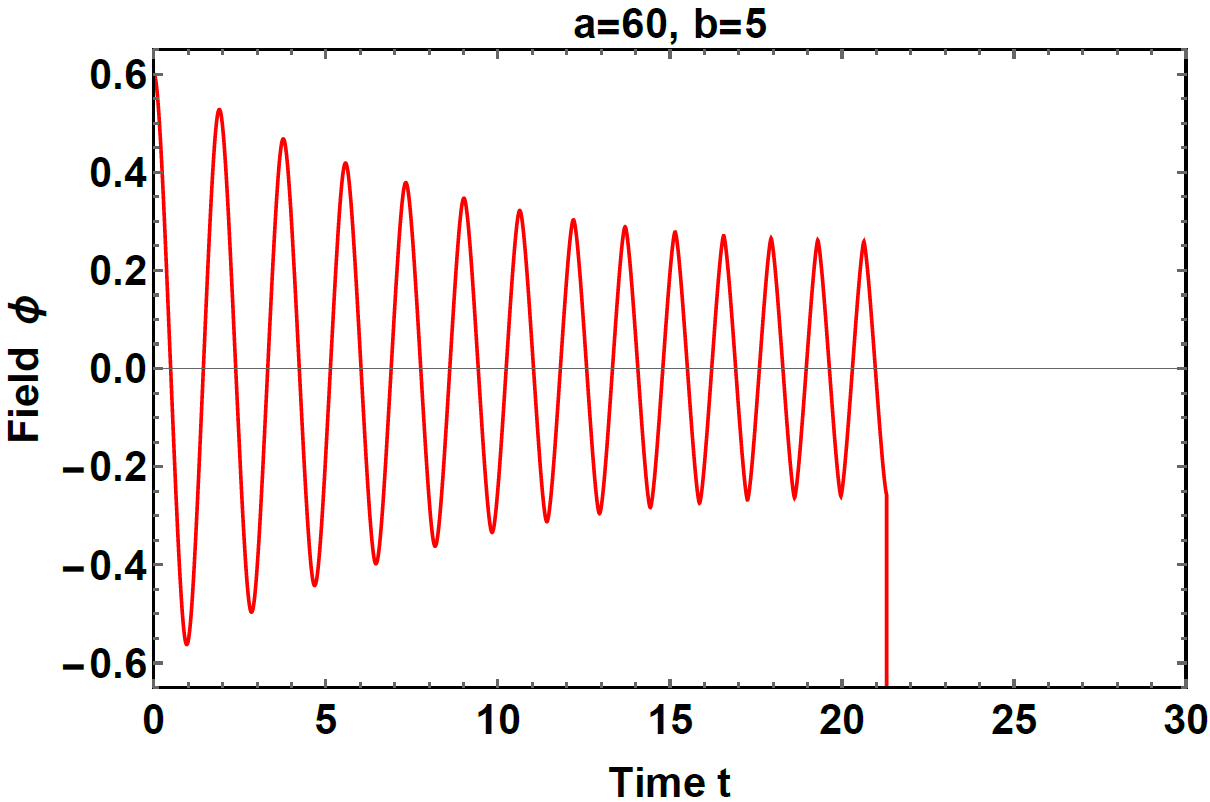}\,\nolinebreak
\includegraphics[scale=0.25]{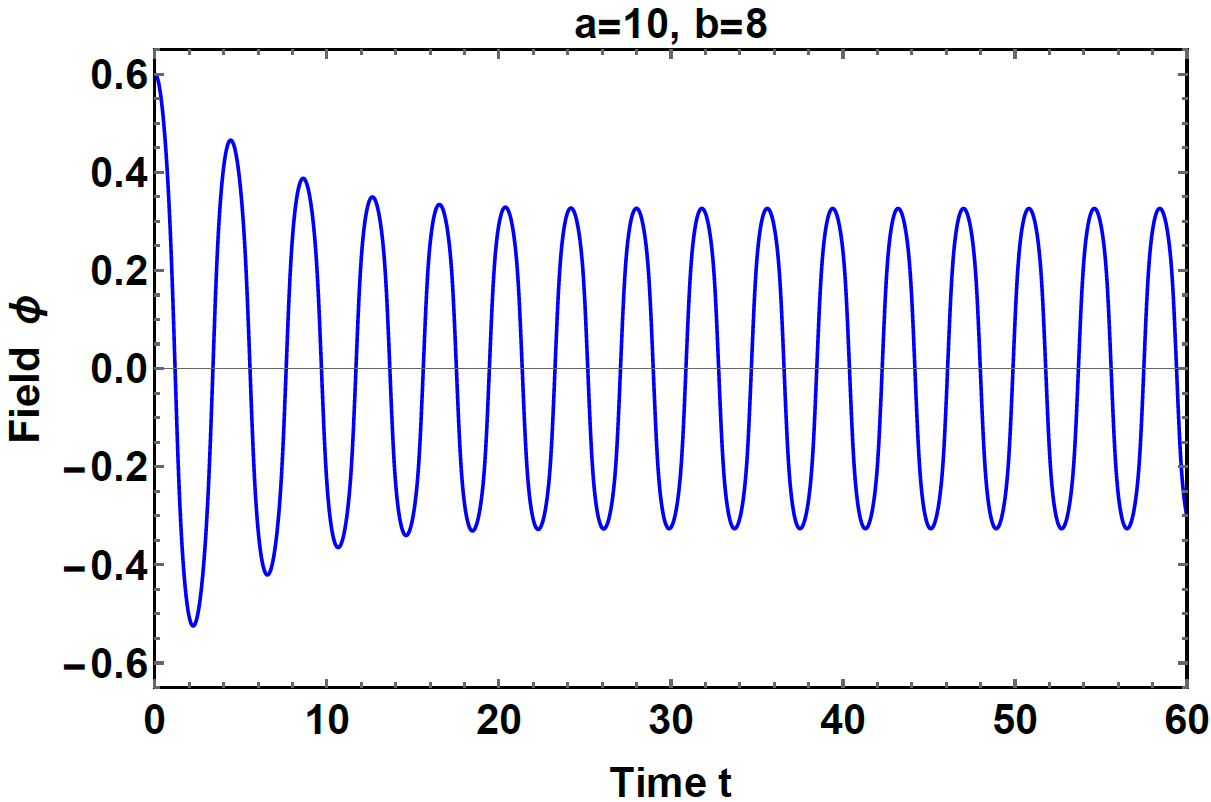}\\
\includegraphics[scale=0.25]{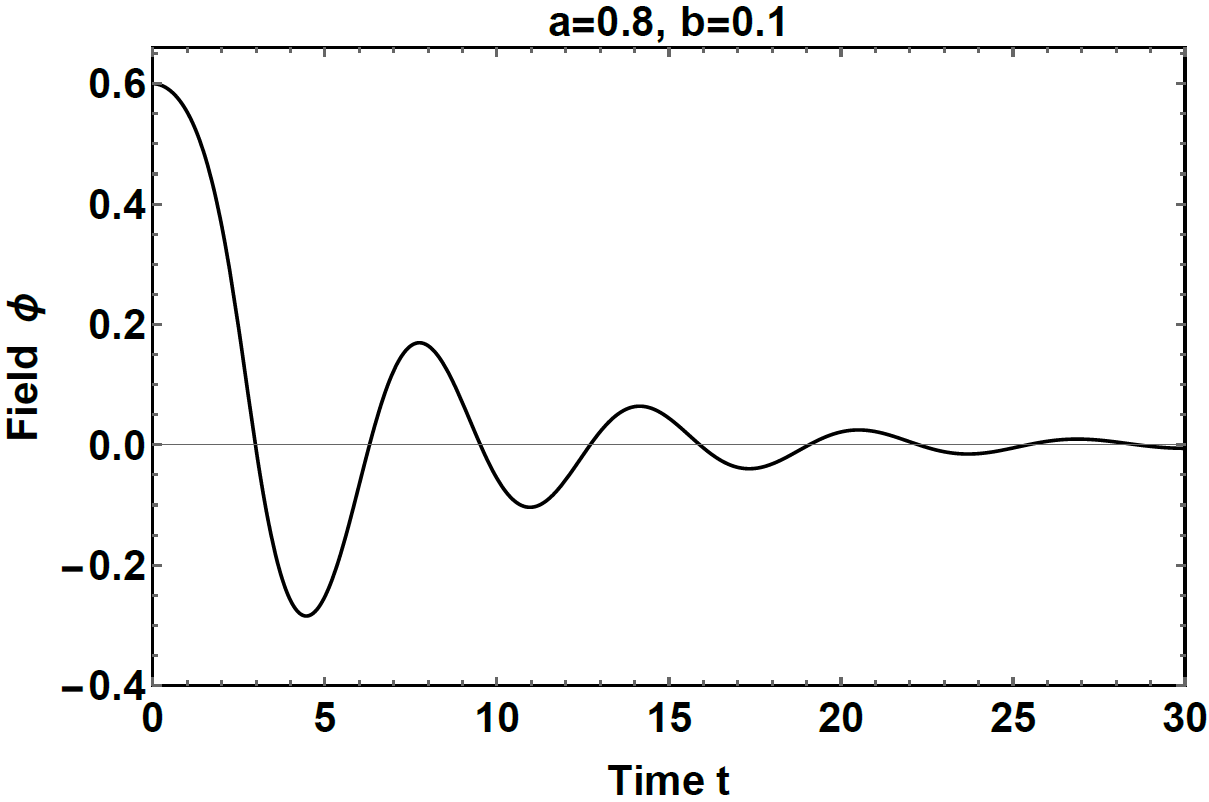}\,\nolinebreak\,
\includegraphics[scale=0.25]{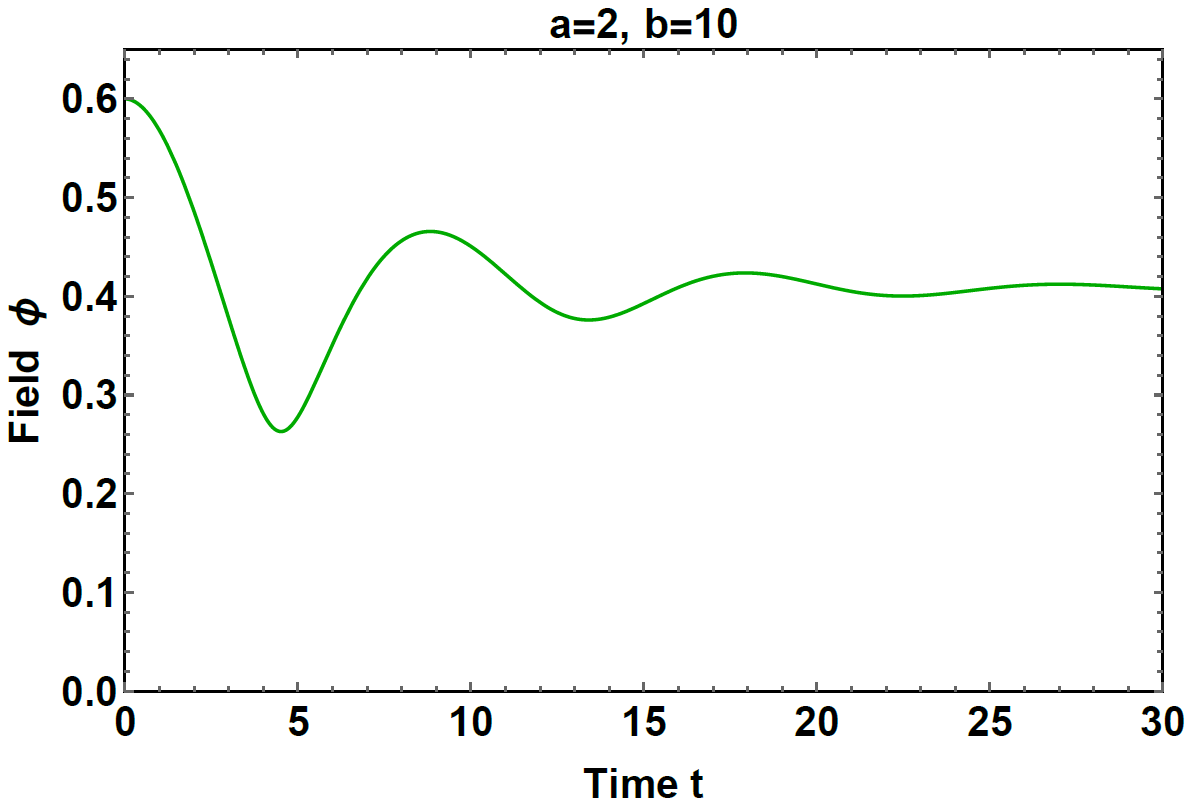}
\caption{Phase diagram as a function of parameters $a$ (vertical axis) and $b$ (horizontal axis) with $\phi_i=0.6, H=0.1$:
(a) upper right blue region corresponds to oscillating solutions, 
(b) lower right green regions corresponds to solutions that approach a constant,
(c) lower left black region corresponds to decaying solutions, 
(d) upper left red region corresponds to singular solutions.
An example of each of these behaviors is plotted in the corresponding four panels.
}\label{solnsample}
\end{figure}

We assume initial conditions
\begin{eqnarray}\label{initial}
\phi(0)&=&\phi_i,\\
\dot{\phi}(0)&=&0
\end{eqnarray}
and a value $H$ for the background Hubble parameter.  
In Figure \ref{solnsample} we pick a particular value of the pair $(\phi_i, H)$, namely $(\phi_i=0.6, H=0.1$), to give a sample of the phase space of solutions.  We display four colors, which represent four distinct asymptotic behaviors, in different regions of the $(a,b)$ plane. They are:

\bigskip

(a) {\em Oscillatory} -- the solution oscillates forever, despite the presence of Hubble friction.

(b) {\em Constant} -- the solution settles down to an extremum of the double well potential. 

(c) {\em Vanishing} -- the solution redshifts towards zero.

(d) {\em Singular} -- at a finite time, the solution hits a singularity in the equation of motion.

In the lower part of Figure \ref{solnsample} we exemplify each of these behaviors, plotting solutions for chosen specific values of the parameters $(a, b)$. 

Among these behaviors the oscillatory is the most novel and surprising, and it will be our main focus. We will describe it in detail in the subsequent sections. In the example plotted in Figure \ref{solnsample} we see the initial amplitude decrease with Hubble friction, then asymptote to a non-zero limiting amplitude for late times. In other cases the amplitude grows before approaching a limiting value asymptotically, as we will discuss.

The constant (fixed point) asymptotic (b)  behavior is unsurprising; it simply involves the field falling to the minimum of the double well potential, which is at
\beq
\phi_0=\pm{1\over\sqrt{3\,a}}.
\eeq

The vanishing asymptotic behavior (c) also has a simple interpretation: it involves the field red-shifting to $\phi\to0$. 

Finally, the singular behavior (d) occurs when
\beq
-1+3\dot\phi^2+3b\phi^2=0,
\eeq
since here the coefficient of $\ddot\phi$ in eq.~(\ref{eomfrac}) becomes zero, as we anticipated earlier.  This possibility is connected to the existence of cusps in the Hamiltonian density.

\subsection{Attractor Behavior}\label{Attract}

\begin{figure}[t]
\centering
\includegraphics[scale=0.65]{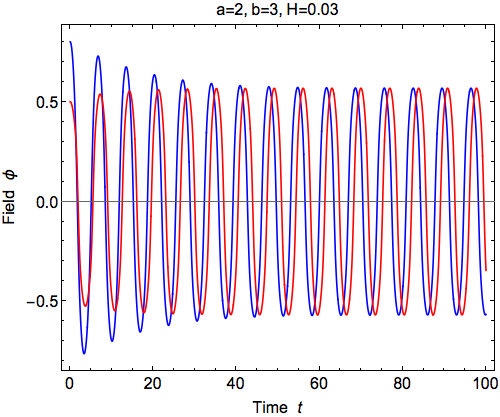}
\caption{Plot of $\phi(t)$ with  $a=2, b=3$, and $H=0.03$. 
The upper blue curve corresponds to a starting value of $\phi_i=0.8$. 
The lower red curve corresponds to a starting value of  $\phi_i=0.5$. Note that Hubble friction causes both to evolve to the same attractor amplitude at late times.}
\label{fig:lowhigh}
\end{figure}

For the oscillatory solutions, Hubble friction drives the solution to an attractor amplitude $\Phi$ at late times. For initial amplitudes a little above or below this attractor amplitude, the field evolves towards it, as seen in Figure \ref{fig:lowhigh}.

In order to understand the condition for the attractor amplitude $\Phi$, let us use the energy density evolution equation
\beq
\dot\rho=-3 H(\rho+p).
\label{rhodot}\eeq
At the attractor we know that the evolution is periodic. Let us time average the solution over a period of oscillation $T_\phi$ and denote time averages by $\langle\ldots\rangle$. The average of some quantity $X$ is given by
\begin{eqnarray}\label{timeavg}
\left<X\right>=\frac{1}{T_{\phi}}\int X dt
=\frac{1}{T_{\phi}}\int \frac{X}{\dot{\phi}}d\phi.
\end{eqnarray}
Conservation of energy gives $\dot\phi$ in terms of $\phi$ and initial amplitude $\phi_i$ as
\begin{eqnarray}\label{phidot}
\dot{\phi}=\sqrt{\frac{-2B+2\sqrt{B^2-3C}}{3}},
\end{eqnarray}
where
\begin{eqnarray}\label{abc}
B&=&\frac{3}{2}b\phi^2-\frac{1}{2}\\
C&=&\frac{3}{4}a\phi^4-\frac{1}{2}\phi^2-\frac{3}{4}a\phi_i^4+\frac{1}{2}\phi_i^2,
\end{eqnarray}
which allows one to evaluate $\langle X\rangle$ as an integral over $\phi$. This includes the integral expression for the period $T_\phi=\int d\phi/\dot\phi$.

Now at the attractor, the average rate of change of energy density should vanish
\beq
\langle\dot\rho\rangle_{att}=0.
\eeq
Hence, using eq.~(\ref{rhodot}), the attractor condition becomes $\langle\rho+p\rangle_{att}=0$. Recalling that the sum of energy density and pressure are related to $\dot\phi$ and the momentum conjugate $\pi_\phi$ by $\rho+p=\pi_\phi\,\dot\phi$, we obtain the condition
\beq
\langle\pi_\phi\,\dot\phi\rangle_{att}=\langle-\dot\phi^2+\dot\phi^4+3\,b\,\phi^2\dot{\phi}^2\rangle_{att}=0.
\label{attractor}\eeq

\subsection{Equation of State}

\begin{figure}[t]
\centering
\includegraphics[scale=0.38]{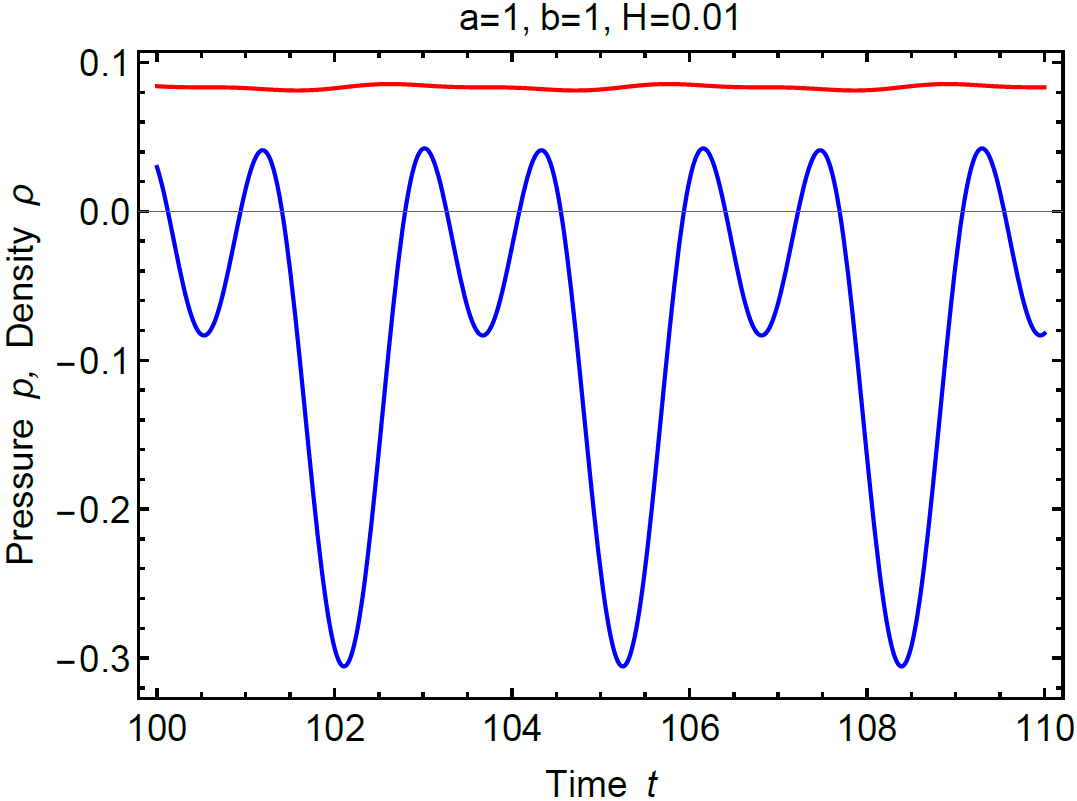}
\caption{The pressure $p$ (lower blue curve) and density $\rho$ (upper red curve) for the late time attractor solution with $a=b=1$ in an expanding FRW universe with $H=0.01$ and $\Lambda=0$. We find that $p$ oscillates significantly from positive to negative values, with mean $\langle p \rangle_{att}=-\langle\rho\rangle_{att}$.} 
\label{eqofstate}
\end{figure}

The instantaneous values of pressure and density are plotted over time in Figure \ref{eqofstate}, where we see significant oscillations in the pressure $p$, and smaller oscillations in the density $\rho$.
Since the standard equations governing expansion in an FRW universe are linear in the density $\rho$ and the pressure $p$, it is appropriate, in assessing the contribution of our oscillatory scalar field, to take time averages over the oscillations, assumed rapid relative to the background expansion.  
As we have just argued, the time average over an oscillation $\langle\rho+p\rangle_{att}$ vanishes.  Therefore, as far as cosmological expansion is concerned, we have an effective equation of state  
\begin{eqnarray}
\left<w\right >_{att} \equiv \left<p\right >_{att} / \left <\rho \right >_{att}=-1.
\end{eqnarray}
This indicates the possibility of de Sitter like solutions. As a possible model of inflation, one would need to find a way to end this phase somehow. For the late-time universe, this would be an interesting form of dark energy. We will return to these issues when we study fluctuations in Section \ref{fluct} and cosmological consequences in Section \ref{Cosmology}.

\section{Special Parameters} \label{ab1} 

In this section we consider the special case in which the parameters in the Lagrangian are $(a,b)=(1,1)$. This leads to tremendous simplifications and enables us to obtain closed-form results. 
By setting $a=1$ and $b=1$ the equation of motion eq.~(\ref{eomfrac})  simplifies to 
\beq
\ddot\phi=-\phi+{3H\dot\phi(1-\dot\phi^2-3\phi^2)\over-1+3\dot\phi^2+3\phi^2}.
\label{ab1eom}\eeq

\subsection{Solution without Hubble Expansion}

Let us first consider the limit in which we ignore Hubble expansion, taking $H\to0$. Then the equation of motion evidently simplifies to just $\ddot\phi=-\phi$, whose
solutions are simply 
\beq\label{cosine}
\phi(t)=\phi_a\cos(t),
\eeq
where $\phi_a$ is a constant amplitude. Although the Lagrangian involves quartic terms that make an $\mathcal{O}(1)$ contribution to the solutions, for the special choice $(a,b)=(1,1)$, the solution, in the absence of Hubble friction, is a pure cosine. We can see why there is something very special about these parameters by considering the energy density. For $(a,b)=(1,1)$, eq.~(\ref{energy}) can be written as a perfect square
\beq
\rho={1\over12}\left(1-3\phi^2-3\dot\phi^2\right)^2+\Lambda.
\label{perfectsquare}\eeq
Since we know that energy is conserved (in the absence of Hubble friction), clearly a pure cosine solution keeps the energy constant. For more general values of $a$ and $b$, $\rho$ is no longer a perfect square, and pure cosines  will no longer provide exact solutions, but we will still find classes of periodic solutions.

The minimum energy configuration occurs at
\beq
\phi_a\bigg{|}_{min,E}={1\over\sqrt{3}},\,\,\,\,\mbox{with}\,\,\,\,\rho_{min}=\Lambda.
\eeq
In the absence of Hubble friction, this leads to perfectly reasonable behavior.  However, it is associated with a cusp of the Hamiltonian, and generic perturbations -- including, as we see in eq.~(\ref{ab1eom}), Hubble friction -- can lead to singularities in the equation of motion.  One might anticipate that friction would drive the system to the minimum energy configuration.  But now we show that Hubble ``friction'' brings in a different attractor.

\subsection{Solution with Hubble Expansion}

When we include a non-zero Hubble parameter $H$ the equation of motion eq.~(\ref{ab1eom}) becomes highly nonlinear. We do not have a closed-form expression for the oscillating solutions, but for small $H$ we can derive a very good approximate solution.

\begin{figure}[t]
\centering
\includegraphics[height=8cm]{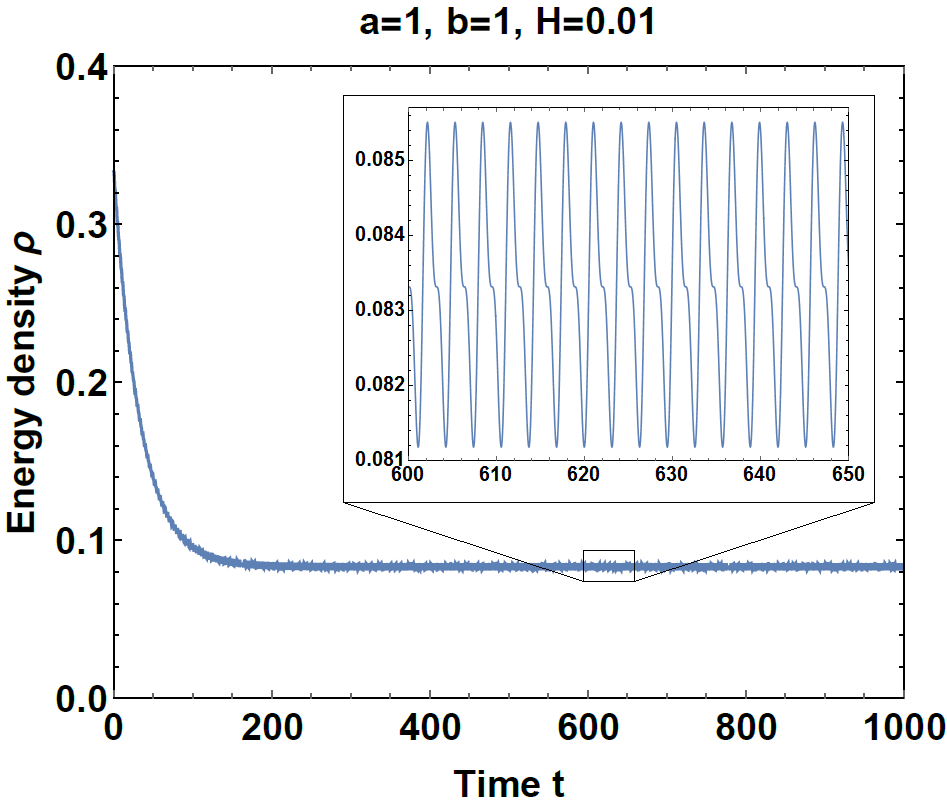}\hspace*{-1.5em}
\caption{Energy density $\rho$ as a function of a time $t$ with $a=b=1$, $\phi_i=1$ and $\Lambda=0$. The outer frame shows the overall behavior. The inner frame shows a zoomed in region of some late time behavior. We see that at late times the energy density is attracted to a non-zero constant, plus very small oscillations.}
\label{rhotime}\end{figure}

For sufficiently small $H$ we expect the solution to remain periodic, as given in eq.~(\ref{cosine}), but we must now allow the amplitude to be a slowly varying function of time. We call the instantaneous amplitude an ``envelope" function $\phi_{env}(t)$ and write 
\beq
\phi(t)=\phi_{env}(t)\cos(t).
\label{envelope}\eeq
To determine the envelope function, we use the conservation of energy equation and takes its time average over a period:
\beq
\langle\dot\rho\rangle=-3H(\langle\rho\rangle+ \langle p\rangle).
\label{timeaveenergy}\eeq
By neglecting the change of $\phi_{env}(t)$ over an oscillation, it is simple to compute the averaged pressure
\beq
\langle p\rangle = -{1\over12}-\Lambda.
\eeq
Substituting this into eq.~(\ref{timeaveenergy}) with constant $H$ allows us to solve for $\langle\rho\rangle$, giving
\beq
\left<\rho\right>=\left(\langle\rho_i\rangle-{1\over12}-\Lambda\right) e^{-3H(t-t_i)}+{1\over12}+\Lambda,
\label{rhoexp}\eeq
where $\rho_i$ is the initial value for the energy density. A plot of the exact numerical result for $\rho(t)$ is given in Figure \ref{rhotime}. We see that there is indeed exponential decay towards a constant. The zoomed in region shows that while oscillations in $\rho$ persist, their amplitudes are very small for $H\ll1$.

\begin{figure}[t]
\centering
\includegraphics[scale=0.4]{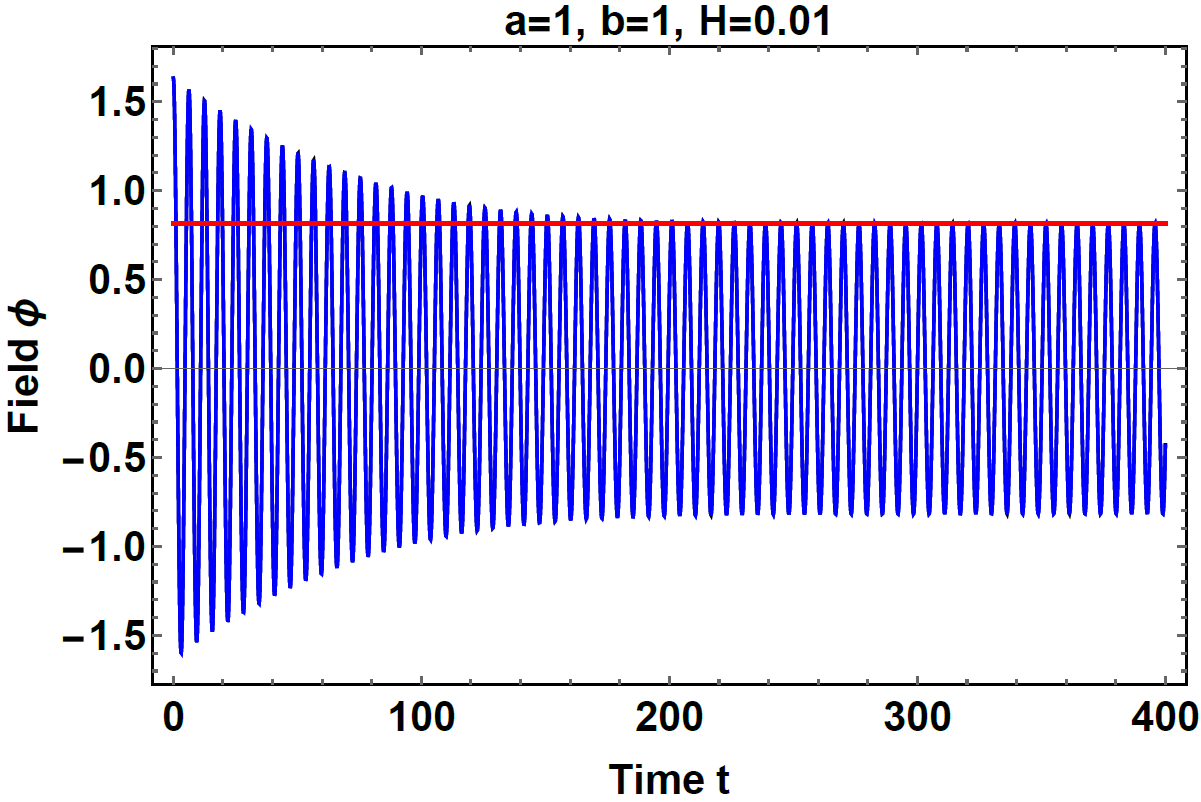}
\caption{The field $\phi$ as a function of time $t$ with $a=b=1$, $H=0.01$, and initial value $\phi_i=2\sqrt{2/3}$. 
The blue curve is the exact numerical $\phi(t)$ reults. 
The black curve is the analytical result given in eqs.~(\ref{envelope},\,\ref{phienv}). The two curves overlap almost exactly.
The horizontal red line is at the attractor value $\Phi_{att}=\sqrt{2/3}$.}
\label{phinumatt}
\end{figure}

With this approximate result for $\langle\rho\rangle$, we can determine the corresponding approximate result for $\phi_{env}(t)$. By substituting eq.~(\ref{envelope}) into eq.~(\ref{perfectsquare}) and neglecting the change of $\phi_{env}(t)$ over an oscillation, we obtain
\beq\label{initialrho}
\left<\rho\right>=\frac{3}{4}\phi_{env}^4-\frac{1}{2}\phi_{env}^2+\frac{1}{12}+\Lambda.
\eeq
By equating this with the exponentially decaying result for $\langle\rho\rangle$ in eq.~(\ref{rhoexp}), we obtain
\beq
\phi_{env}(t)=\sqrt{\frac{1+\sqrt{1+3\phi_i^2(-2+3\phi_i^2)e^{-3H(t-t_i)}}}{3}}.
\label{phienv}\eeq
Figure \ref{phinumatt} compares this analytical result to the exact numerical result. There is remarkably good agreement. 

For late times, eq.~(\ref{phienv}) predicts that for any initial starting value $\phi_i$ and for any $H$, the envelope function asymptotes to the following attractor amplitude with corresponding attractor energy density:
\bea
\phi_{a,att}=\Phi_{att}\amp=\amp\sqrt{2\over3},\\
\langle\rho\rangle_{att}\amp=\amp{1\over12}+\Lambda.
\eea
Note that $\langle\rho\rangle_{att}>0$ even if $\Lambda=0$, so the dynamics of $\phi$ is generating a non-zero energy density.
Earlier we explained that the attractor behavior is determined by the condition that $\langle\rho+p\rangle=0$, which in this class of models was written in eq.~(\ref{attractor}). By demanding this be satisfied, we  again see $\Phi_{att}=\sqrt{2/3}$. 

Hence we see that for generic initial conditions, the system evolves towards an attractor solution.  In that solution the field oscillates with finite amplitude and non-zero energy density. The oscillations in $\phi$ are large, but the oscillations in $\rho$ are small.   This behavior permits an approximate de Sitter phase, as we will elaborate further in Section \ref{Cosmology}.

\section{General Parameters}\label{closeto11} 

As we summarized in Section \ref{overview}, the solutions of the model exhibit a variety of behaviors depending on the values of $(a, b, \phi_i, H)$.  However, in the previous section we saw that the behavior is always oscillatory when $(a,b)=(1,1)$. In this Section, we begin by studying $(a,b)$ close to $(1,1)$, focussing on the oscillatory solutions, and then we explore general values for $(a,b)$ and describe the full phase space of solutions.

\subsection{Quasi-Special Parameters}\label{Quasi} 

Here we suppose that both $a$ and $b$ are not far from 1, which will allow us to obtain a solution perturbatively. Let us parameterize the departure from those values as
\beq
a=1+\alpha,\,\,\,\,\,\,b=1+\beta,
\label{abbeta}\eeq
with $|\alpha|\ll 1$, $|\beta|\ll1$, and $\alpha,\,\beta$ allowed to be positive or negative.

We will construct a perturbative solution in powers of $\alpha$ and $\beta$. In order to have a {\em single} expansion parameter, we consider that every power of $\alpha$ to be of the same order as the corresponding power of $\beta$. To make this explicit, we will rewrite $\alpha=\aaa\,\beta$, treating $\aaa$ as $\mathcal{O}(1)$, thereby leaving $\beta$ as the residual expansion parameter. 

For $a$ and $b$ different from 1, the frequency of oscillations will be shifted away from 1.
So we expand both the field and the frequency as a power series in $\beta$ as follows
\begin{eqnarray}\label{boxedbeta2}
\phi(\tau)&=&\phi_0(\tau)+\phi_1(\tau)\,\beta+\phi_2(\tau)\label{phi}\,\beta^2+\dots\label{phiexp}\\
\omega&=&\sqrt{1+c_1\,\beta+c_2\,\beta^2+\dots}, \label{omegaexp}
\end{eqnarray}
where the new time variable is $\tau=\omega\, t$ and the $c_i$ are constants to be determined.

\subsubsection{Perturbative Solution without Hubble Expansion}\label{pertnohubble}

In the absence of Hubble friction ($H=0$), the equation of motion in terms of $a$, $b$, and $\omega$ is
\beq\label{eomfractau}
\phi''=\frac{\phi\left(1-3a\phi^2-3b\omega^2\phi'^2\right)}{-\omega^2+3\omega^4\phi'^2+3b\omega^2\phi^2},
\eeq
where primes $'$ indicate derivatives with respect to $\tau$. We substitute eqs.~(\ref{abbeta},\,\ref{phiexp},\,\ref{omegaexp}) into this and work order by order in powers of $\beta$.

At $\mathcal{O}(\beta^0)$ the equation of motion is obviously $\phi_0''+\phi_0=0$, with solution
\beq
\phi_0(\tau)=\phi_a\cos(\tau),
\eeq
for any constant $\phi_a$.  

At $\mathcal{O}(\beta)$ the equation of motion is
\beq
\phi_1''+\phi_1=
\frac{\left(4c_1\phi_a-3(2+4c_1-3\aaa)\phi_a^3\right)\cos(\tau)+3(-2+\aaa)\phi_a^3\cos(3\tau)}{4\left(1-3\phi_a^2\right)}.
\label{phi1eom}\eeq
Note that the terms on the right hand side include a term that is proportional to $\cos(\tau)$. If this term is present it will act as a driving term that is on resonance with the natural frequency of oscillation on the left hand side. That would lead to secular growth in $\phi_1$ over time, rather than periodicity. Hence, we require the coefficient of $\cos(\tau)$ to vanish, which forces $c_1$ to take the value
\begin{eqnarray}\label{omgconst1}
c_1=\frac{3(-2+3\aaa)\phi_a^2}{-4(1-3\phi_a^2)}.
\end{eqnarray}
Upon substitution of this into eq.~(\ref{phi1eom}) and focussing on the particular solution, we obtain
\beq\label{phi1}
\phi_1(\tau)=\frac{3(-2+\aaa)\phi_a^3}{32(-1+3\phi_a^2)}\cos(3\tau).
\eeq

At $O(\beta^2)$ it is a straightforward procedure to obtain the equation of motion, then obtain $c_2$ to avoid secular behavior and then determine $\phi_2(\tau)$. The results are somewhat unwieldy and we report on them in Appendix \ref{Appbeta2}.

We note that these higher order terms, $\phi_1,\,\phi_2,\ldots$ cause a shift in the amplitude of oscillation away from $\phi_a$. 
Evaluating $\phi(\tau)$ at $\tau=0$, we get a relation between the amplitude $\Phi=\phi_{max}$ and the coefficient of the cosine $\phi_a$ as follows
\beq
\Phi=\phi_a+\frac{3(-2+\aaa)\phi_a^3}{32(-1+3\phi_a^2)}\,\beta+\Phi_2\,\beta^2+\dots,
\eeq
where $\Phi_2$ is again given in Appendix \ref{Appbeta2}.

\begin{figure}[t]
\label{earlylate1}
\centering
\includegraphics[scale=0.25]{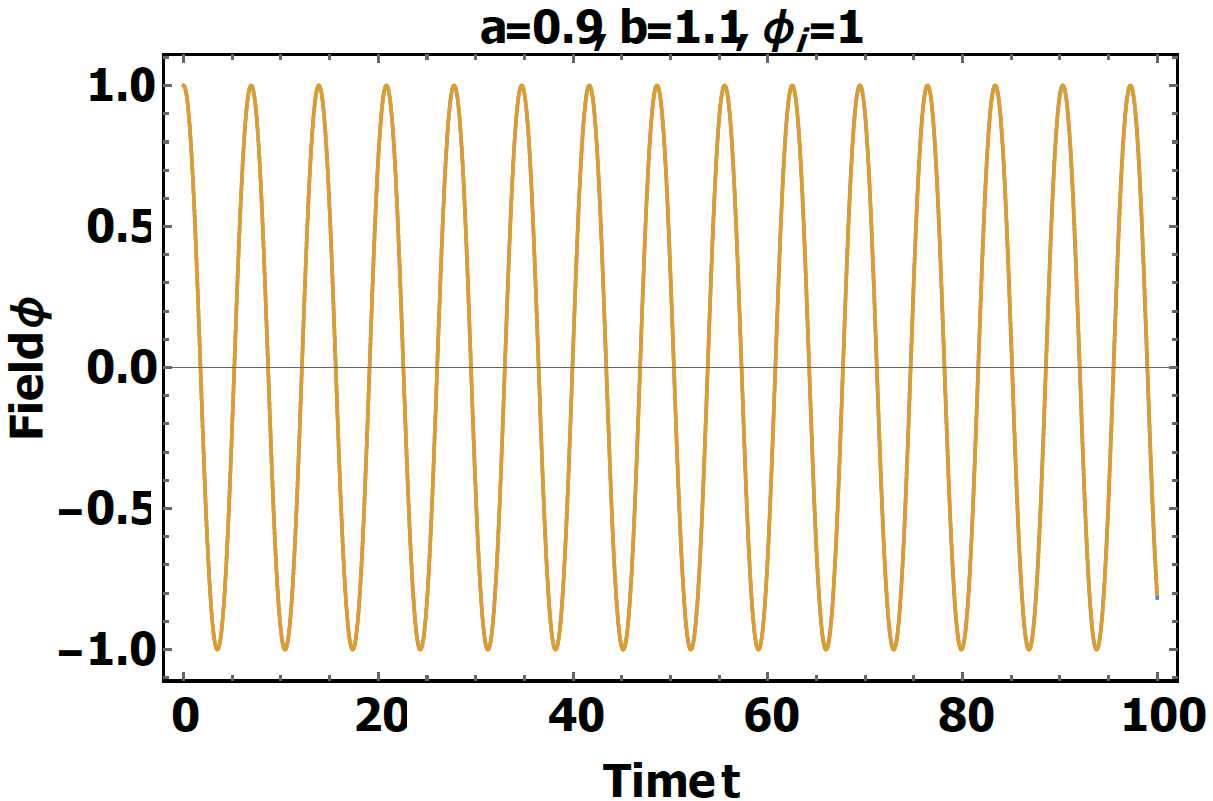}
\includegraphics[scale=0.25]{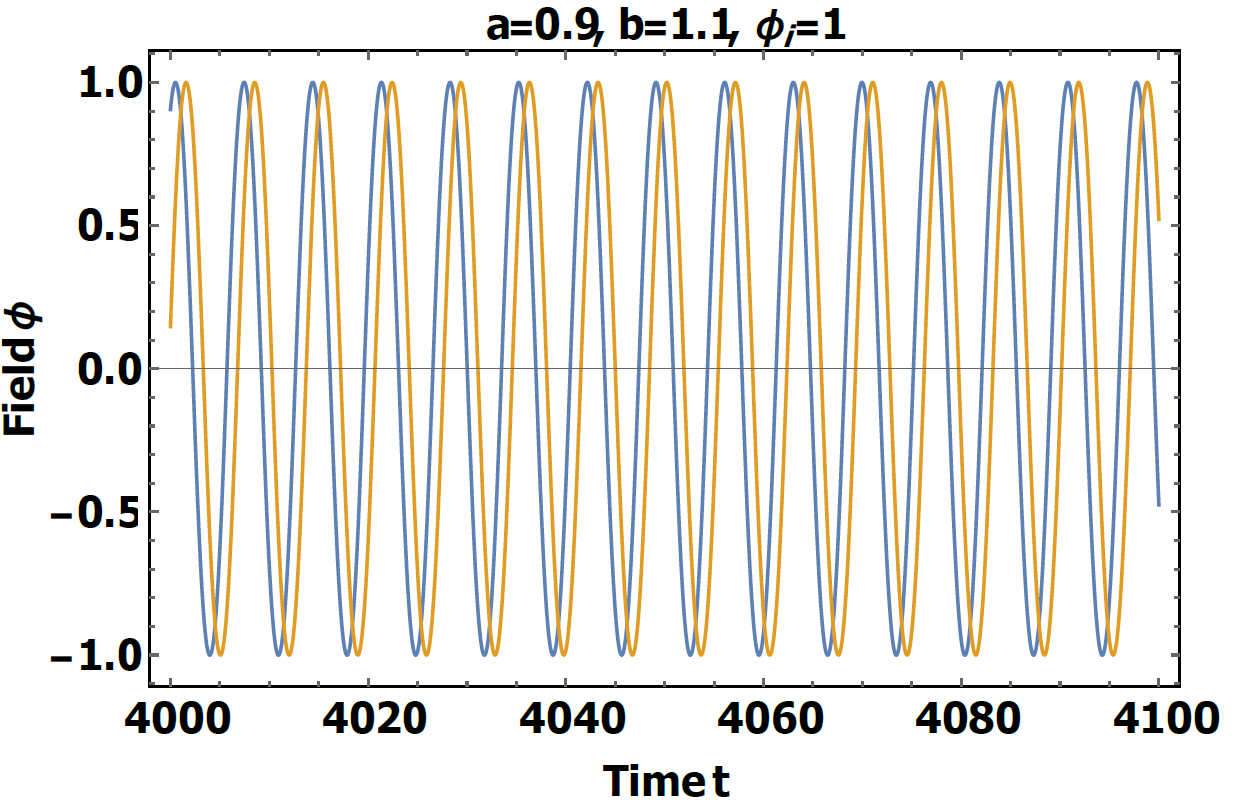}
\caption{Comparative plots of the field $\phi$ as a function of time $t$, with parameters $a=0.9, b=1.1, \phi_i=1$ with $H=0$. The blue curve is the analytical solution $\phi(t)$ computed to $\mathcal{O}(\beta^2)$. The orange curve is the exact numerical solution $\phi_{num}(t)$. The left panel is for early times. The right panel is for later times.}
\label{earlylate2}
\end{figure}

A plot of the early and late time solution is shown in Figure \ref{earlylate2}, where we compare the exact numerical result to the analytical result computed to $\mathcal{O}(\beta^2)$, for parameters $a=0.9, b=1.1, \phi_i=1$ and $H=0$. We see remarkably good agreement, with only a small phase shift noticeable at late times, reflecting higher order corrections coming in at $\mathcal{O}(\beta^3)$.

\subsubsection{Perturbative Solution at Attractor}

Upon inclusion of Hubble friction, the solution settles down to an attractor amplitude $\Phi$ corresponding to an attractor value of $\phi_a$.
At the attractor, we perform an expansion of $\phi_a$ in powers of $\beta$
\begin{eqnarray}
\phi_{a,att}=\zeta_0+\zeta_1\beta+\zeta_2\beta^2+\dots.
\end{eqnarray}
Substituting this into the attractor condition eq.~(\ref{attractor}) along with the solution for $\phi(\tau)$ determined above, and working to $\mathcal{O}(\beta^2)$, we find
\begin{eqnarray}\label{phiaatt}
\phi_{a,att}=\sqrt{\frac{2}{3}}-\frac{1}{4}\sqrt{\frac{3}{2}}\,\aaa\,\beta+\frac{7(-4+4\aaa+19\aaa^2)}{256\sqrt{6}}\beta^2+\dots.
\end{eqnarray}
We then obtain $\phi(\tau)$ at the attractor amplitude as
\begin{eqnarray}
\phi(\tau)_{att}=\sqrt{\frac{2}{3}}\cos(\tau)+\frac{-6\,\aaa\cos(\tau)+(-2+\aaa)\cos(3\tau)}{8\sqrt{6}}\beta +\phi_2(\tau)_{att}\beta^2+\dots,
\end{eqnarray}
where the $O(\beta^2)$ result $\phi_2(\tau)_{att}$ is reported in Appendix \ref{Appbeta2}. The corresponding frequency is found to be
\begin{eqnarray}
\omega^2\vert_{att}=1+\left(-1+\frac{3}{2}\aaa\right)\beta+\frac{1}{32}\left(20-28\,\aaa+\aaa^2\right)\beta^2+\dots.
\end{eqnarray}
Comparing this analytical result at the attractor to the exact numerical result, we again find excellent agreement.

\subsection{Arbitrary Parameters}\label{generalab}

\subsubsection{Phase Diagrams}

In this section, we scan through arbitrary values of the parameters $a$ and $b$. For generic values of $a$ and $b$ (far away from $(1,1)$) we do not have a good analytical handle on the general form of the solutions, but we can solve the equations numerically.  Our results are captured in the color-
\begin{figure}[H]
\centering
\includegraphics[scale=0.37]{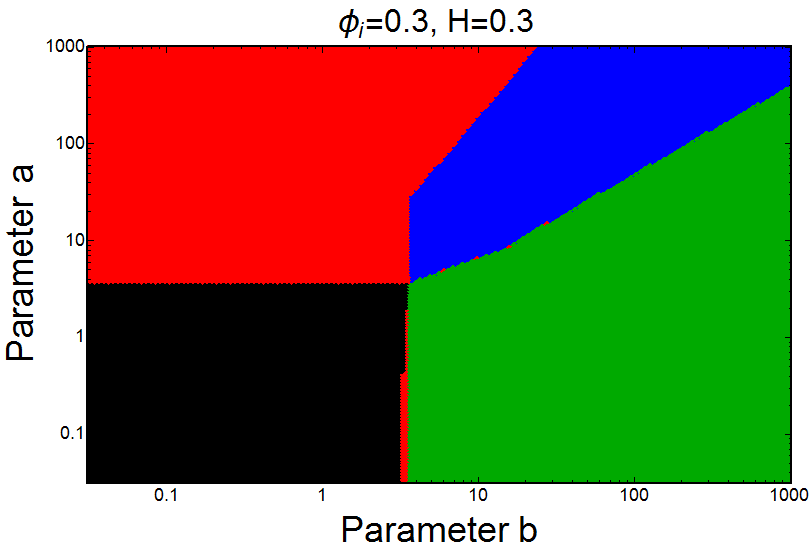}
\includegraphics[scale=0.37]{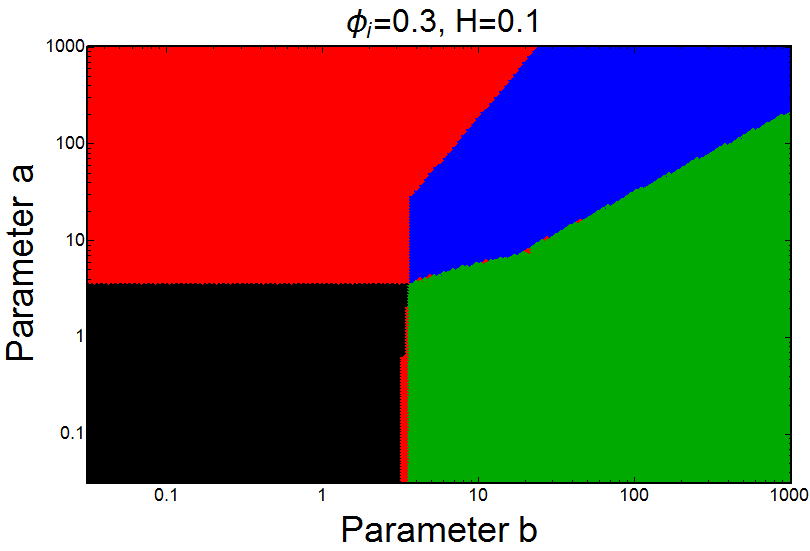}
\includegraphics[scale=0.37]{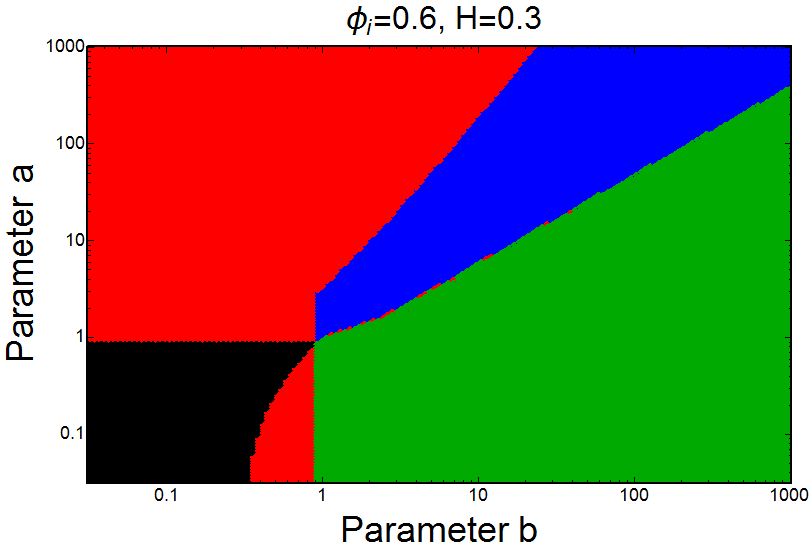}
\includegraphics[scale=0.37]{c=0point6H=0point1nolines.png}
\includegraphics[scale=0.37]{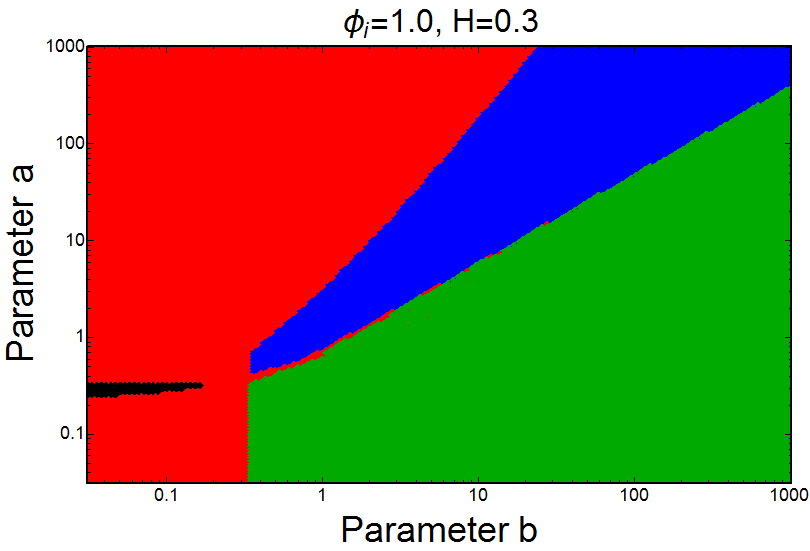}
\includegraphics[scale=0.37]{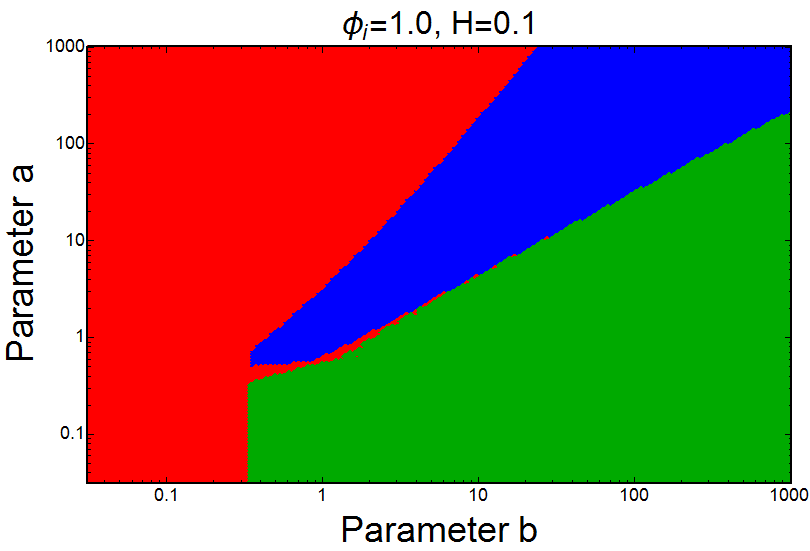}
\caption{Phase diagrams as a function of parameters $a$ (vertical axis) and $b$ (horizontal axis). Each color represents a different behavior from numerically solving for some initial condition $\phi_i$ and Hubble $H$.}
\label{bigphaseNew}
\end{figure}
\hspace{-0.6cm}coded ``phase diagram'' Figure \ref{bigphaseNew}, where here, as earlier, we encode asymptotic behavior which is
oscillatory as blue (a), constant as green (b), vanishing as black (c) and singular as red (d).

An obvious feature is that as we increase Hubble friction, the size of the black region describing vanishing solutions that redshift to zero, increases. We also find that as we increase the initial starting amplitude $\phi_i$ the phase diagram becomes more universal.

\begin{figure}[t]
\centering
\includegraphics[scale=0.4]{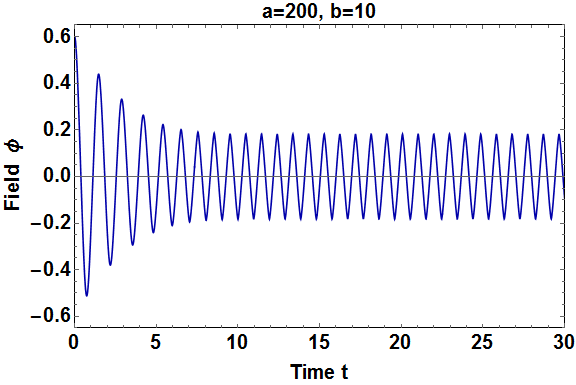}
\includegraphics[scale=0.4]{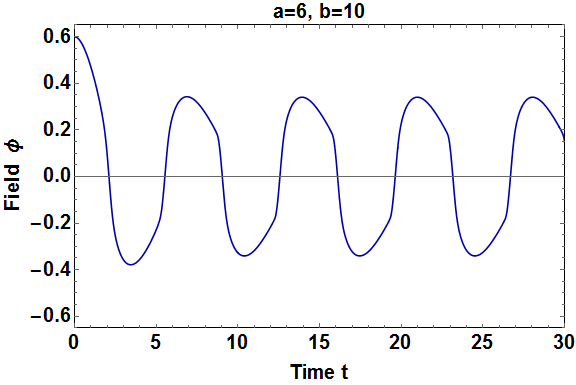}
\includegraphics[scale=0.4]{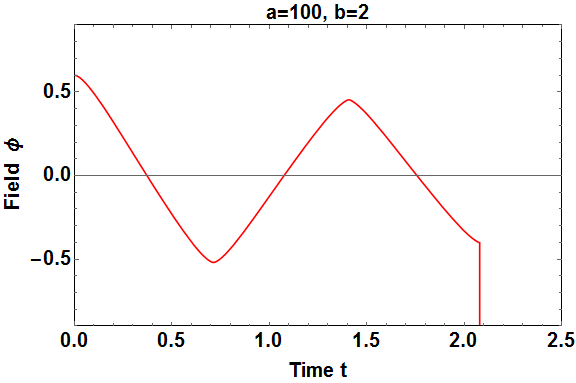}
\includegraphics[scale=0.4]{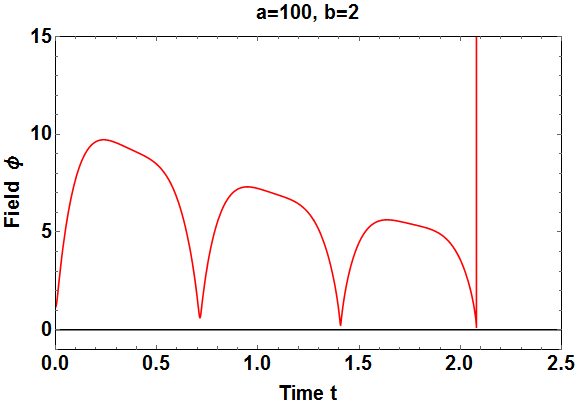}
\caption{Plots of the exact numerical result $\phi_{num}(t)$ with $\phi_i=0.6$ and $H=0.3$ for a selection of different $a$ and $b$ values. 
Top left panel: sawtooth solution found near blue/red boundary with $a=200$ and $b=10$. 
Top right panel: semi-circular solution found near blue/green boundary with $a=6$ and $b=10$. 
Bottom panel: singular solution with $a=100$ and $b=2$; bottom left is $\phi_{num}(t)$; bottom right is the corresponding value of the coefficient of $\ddot\phi$ from eq.~(\ref{eomfrac}).}
\label{blowupdenom}
\end{figure}	

Earlier we showed that for $(a,b)$ near $(1,1)$ the oscillatory solution is almost harmonic (and it is exactly harmonic for $a=b=1$). However, for values of $a$ and $b$ far away from these special values, we can find rather different oscillatory behavior.
In particular, when we are near the blue/red boundary, we find the solutions look closer to a sawtooth, whilst when we are near the red/blue boundary, we find the solutions look closer to a series of alternating semi-circles. Some representative examples of their shapes are given in the top row of Figure \ref{blowupdenom}. 
Also, we show some features of a red singular solution whereby the coefficient of $\ddot\phi$ in eq.~(\ref{eomfrac}) vanishes, as plotted in the bottom row of Figure \ref{blowupdenom}.

\subsubsection{Phase Boundaries}

In this section, we provide analytical derivations of the four main boundaries in the phase diagram. We are able to derive two of the boundaries exactly, whilst providing only rough estimates for the two other boundaries. 
The two boundaries that we can derive exactly are:
\begin{itemize}
\item The vertical boundary separating the (green) constant phase from the (red) singular phase. This can be obtained by finding when the coefficient of $\ddot\phi$ is zero at the initial time (where $\dot\phi_i=0$), yielding
\beq
b={1\over 3\phi_i^2}.
\eeq
\item The horizontal boundary separating the (black) vanishing phase from the (red) singular phase. This can be obtained by finding when right hand side of eq.~(\ref{eomfrac}) is zero at the initial time (where $\dot\phi_i=0$), yielding
\beq
a={1\over 3\phi_i^2}.
\eeq
\end{itemize}

\begin{figure}[t]
\centering
\includegraphics[scale=0.4]{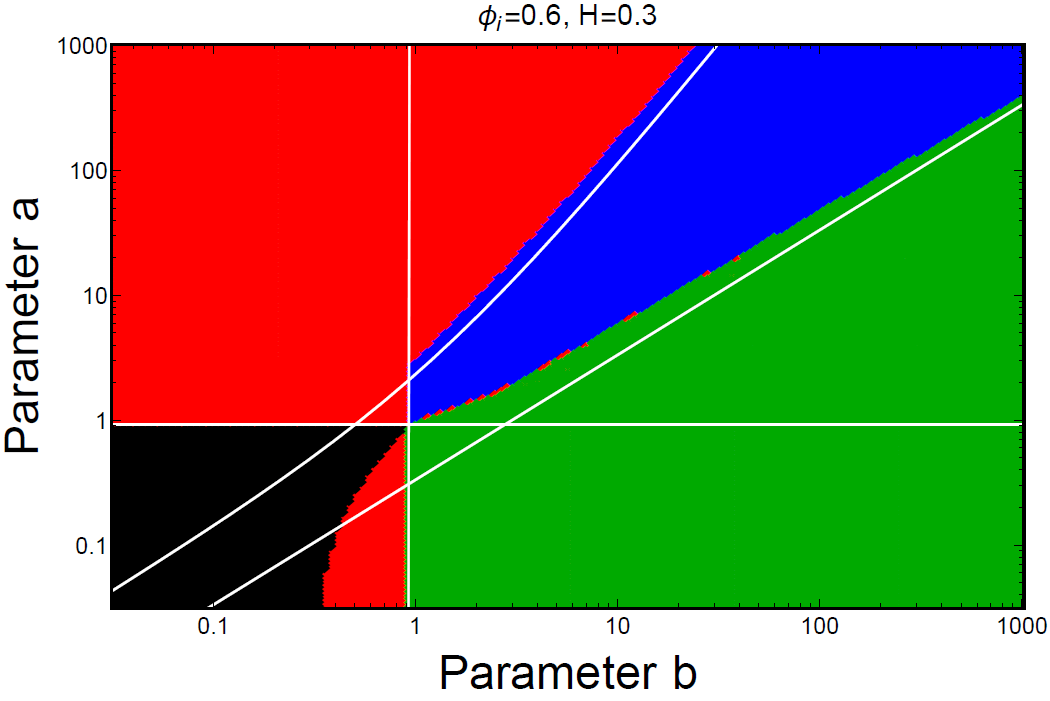}
\caption{An example phase diagram with derived boundary curves in white for $\phi_i=0.6$ and $H=0.3$. The horizontal and vertical boundaries are exact derivations, whereas the diagonal boundaries were derived using very rough single harmonic approximations.}
\label{boundfig}
\end{figure}

We now derive semiquantitative estimates for the two boundaries surrounding the oscillatory region. To do so, we make the very rough approximation that the oscillatory solutions are governed by a single harmonic, such as $\phi(t)\approx\phi_a\,\cos(\omega\,t)$ for some amplitude $\phi_a$ and frequency $\omega$. As shown earlier, for $a$ and $b$ far from $(1,1)$ this is a rather poor approximation, but we will use it here to get a rough idea of the boundaries. Inserting this into (\ref{eomfrac}) with $H=0$ and keeping only terms proportional to the leading harmonic leads to the relationship
\begin{eqnarray}
\phi_a^2\approx\frac{4}{3}\frac{1-\omega^2}{\left(3a-\omega^2(2b+\omega^2)\right)}.
\end{eqnarray}
Then demanding that we are at the attractor solution $\langle\pi_\phi\,\dot\phi\rangle=0$ gives a second relationship
\begin{eqnarray}
\phi_a^2\approx\frac{4}{3(\omega^2+b)}.
\end{eqnarray}
The two boundaries that we can then estimate are:
\begin{itemize}
\item The boundary separating the (blue) oscillating phase from the (green) constant phase.
This can be obtained by finding when $\omega=0$ in the above pair of equations, yielding
\begin{eqnarray}
a\approx\frac{1}{3}b.
\end{eqnarray}
\item The boundary separating the (blue) oscillating phase from the (red) singular phase.
This can be obtained by finding when the coefficient of $\ddot\phi$ from eq.~(\ref{eomfrac}) touches zero using the above estimates, yielding
\begin{eqnarray}
a\approx\frac{1}{3}b(4+3b).
\end{eqnarray}
\end{itemize}
An example of the boundary fits is shown in Figure \ref{boundfig}.

\section{Fluctuations}\label{fluct}

Previously we examined a classical homogeneous background $\phi=\phi(t)$, and found a large parameter regime of oscillatory solutions in an expanding universe, which spontaneously break time translation symmetry. In this section we include quantum fluctuations in the field.

Lets promote the field to a quantum operator in the Heisenberg picture as
\beq
\hat\phi(t,{\bf x})=\phi(t)+\hat{\delta\phi}(t,{\bf x}),
\label{heisenberg}\eeq
where $\phi(t)$ is the classical homogeneous background we examined in the previous sections and $\hat{\delta\phi}(t,{\bf x})$ is a quantum fluctuation that can depend on space and time. Here we will only work to linear order in the quantum fluctuations $\hat{\delta\phi}$. At the linearised level, the solution can be decomposed in terms of mode functions $v_k(t)$ as
\beq
\hat{\delta\phi}(t,{\bf x})=\int\! {d^3k\over(2\pi)^3}\left[\hat{a}_k\,v_k(t)\,e^{i{\bf k}\cdot{\bf x}}+\hat{a}_k^\dagger\,v_k^*(t)\,e^{-i{\bf k}\cdot{\bf x}}\right],
\eeq
where $\hat{a}_k^\dagger$ and $\hat{a}_k$ are standard creation and annihilation operators.

We substitute $\hat\phi$ into the Heisenberg equation of motion and linearize to obtain the following equation for the mode functions
\beq\label{perteom}
d_1(t)\ddot{v}_k(t)+d_2(t)\dot{v}_k(t)+d_3(t)v_k(t)=0,
\eeq
where the coefficients are
\bea
d_1(t)\amp=\amp1-3\dot{\phi}^2-3b\phi^2,\\
d_2(t)\amp=\amp-6\,\dot\phi(\ddot{\phi}+b\,\phi),\\
d_3(t)\amp=\amp (1-\dot{\phi}^2-3b\phi^2)k^2+1-6b\ddot{\phi}\phi-3b\dot{\phi}^2-9a\phi^2,
\eea
and we ignore corrections from Hubble friction.

In the oscillatory regime, $\phi$ is periodic and so the coefficient functions $d_1,\,d_2,\,d_3$ are all periodic too. Then by Floquet's theorem, the solutions are of the form
\begin{eqnarray}\label{hill}
v_k(t)=P_1(t)e^{\mu_k t}+P_2(t)e^{-\mu_k t},
\end{eqnarray}
where $\mu_k$ is a complex number called the \emph{Floquet exponent} and $P_1(t)$ and $P_2(t)$ are periodic functions. In general, $\mu_k$ depends on the wavenumber, with $Re(\mu)\neq0$ corresponding to exponential growth and $Re(\mu)=0$ corresponding to purely oscillatory evolution. The Floquet exponents evidently determine the late time behavior of the system.

\subsection{Special Parameters}

For $a=b=1$, we earlier showed that we have the exact oscillatory background solution (ignoring corrections from $H$) as $\phi(t)=\phi_a\cos(t)$. Here $\phi_a=\sqrt{2/3}$ at the attractor solution, but is otherwise arbitrary. In this case the equation of motion for the mode functions (\ref{perteom}) simplifies considerably. We find that the resulting equation is of the form of the Mathieu equation
\begin{eqnarray}
\ddot{v}_k+\left[\tilde{A}-2\,\tilde{q}\cos(2\,t)\right]\!v_k=0
\end{eqnarray}
with
\begin{eqnarray}
\tilde{A}\amp=\amp1+\frac{k^2(1-2\phi_a^2)}{1-3\phi_a^2},\\
\tilde{q}\amp=\amp\frac{k^2\phi_a^2}{2(1-3\phi_a^2)}.
\end{eqnarray}
The Mathieu equation has known numerical solutions for the Floquet exponent $\mu_k$. The method to determine such exponents will be reviewed in Section \ref{floquetsection}.

\begin{figure}[t]
\centerline{\includegraphics[scale=0.55]{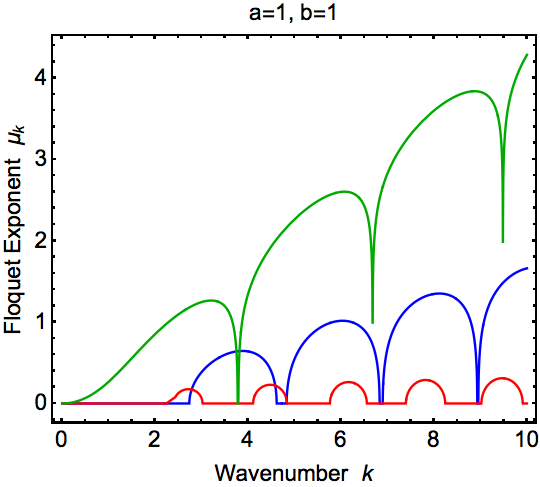}}
\caption{Magnitude of real part of Mathieu exponent $|\mathbb{R}(\mu_k)|$ as a function of wavenumber $k$ for different amplitudes $\phi_a$ ($a=b=1$). The upper (green) curve is for $\phi_a=0.85\sqrt{2/3}$. The middle (blue) curve is for $\phi_a=\sqrt{2/3}$. The lower (red) curve is for $\phi_a=1.2\sqrt{2/3}$.}
\label{floqx3}
\end{figure}
In Figure \ref{floqx3} we plot the magnitude of the real part of $\mu_k$ as a function of wavenumber $k$ for different amplitudes $\phi_a$. We see in the upper (green) curve that for $\phi_a<\phi_{a,att}=\sqrt{2/3}$ there is considerable instability, even extending down to low wavenumbers. On the other hand, we see in the middle (blue) curve that for $\phi_a=\phi_{a,att}=\sqrt{2/3}$ there is still large instability at high wave numbers, but no instability for $k\lesssim 2$. Futhermore, we see in the lower (red) curve that for $\phi_a>\phi_{a,att}=\sqrt{2/3}$ the instability bands continue to get thinner and weaker.

\subsection{Arbitrary Parameters}

\subsubsection{Floquet Theory}\label{floquetsection}

Here we elaborate on the numerical method to determine the Floquet exponents $\mu_k$ by reducing the calculation to a $2\times2$ matrix problem.

To form a complete basis of solutions, we evolve the following set of orthonormal initial conditions through one period $T$ of the background
\begin{eqnarray}
 \left( \begin{array}{c}
v_k(0)  \\
\dot{v}_k(0)  \end{array}\right)
\equiv
 \left( \begin{array}{c}
1  \\
0  \end{array}\right)\rightarrow
 \left( \begin{array}{c}
v_k(T)  \\
\dot{v}_k(T)  \end{array}\right)_1,\,\,\,\,\,\,
 \left( \begin{array}{c}
v_k(0)  \\
\dot{v}_k(0)  \end{array}\right)
\equiv
 \left( \begin{array}{c}
0  \\
1  \end{array}\right)\rightarrow
 \left( \begin{array}{c}
v_k(T)  \\
\dot{v}_k(T)  \end{array}\right)_2 .
\end{eqnarray}
We then form the matrix $\bold{M}(T)$ built out of the column vectors through one cycle $()_1$ and $()_2$.
Since the coefficients in the differential equation are periodic, the evolution through $N$ periods is 
\begin{eqnarray}
\bold{M}\left(N\,T\right)=\bold{M}(T)^N.
\end{eqnarray}
We diagonalize the matrix as
\begin{eqnarray}
\bold{M}(N\,T)=\bold{A}
\left( \begin{array}{cc}
\lambda_1^N & 0  \\
0 & \lambda_2^N  \end{array} \right)
\bold{A}^{-1},
\end{eqnarray}
where $\lambda_1$, $\lambda_2$ and $\bold{A}$ are the eigenvalues and matrix of eigenvectors of $\bold{M}$, respectively. 
The Floquet exponent $\mu_k$, defined through $\sim e^{\pm\mu_k t}$, is therefore given by
\begin{eqnarray}
\mu_{k}=\frac{1}{T}\log(\lambda_i).
\end{eqnarray}
Using the numerically found oscillatory solution for the background $\phi(t)$, we can insert into (\ref{perteom}), evolve through one cycle, and determine the Floquet exponents for any appropriate choice of parameters.

\subsubsection{Contour Plots}

In Figure \ref{contourplots}, we show contour plots of the Floquet exponent as a function of wavenumber $k$ and parameter $b$, with $a$ fixed at several values. For each $a$ and $b$ value, we have allowed the background to relax to its attractor amplitude, and then studied fluctuations around it. The range for $b$ plotted is the {\em entire} range that is compatible with oscillatory solutions for each corresponding $a$ value. 

We observe significant instability bands for high wavenumbers. We also note that a much weaker band can exist for low $k$ depending sensitively on the $a$ and $b$ values. We note for a fixed $a$, there is a large range of $b$ values, in the upper part, that are stable for small wavenumbers.

\subsection{Small Wavenumbers}

The existence of stability for long wavelengths for a range of $a$ and $b$ values is especially interesting. If the higher wave numbers are beyond the cutoff of the effective field theory, then only the small wavenumber regime is relevant. This may allow the oscillatory phase to last indefinitely. Here we investigate  the small $k$ regime in more detail.

\subsubsection{Floquet Squared without Hubble Expansion}\label{floquetnohub}

For precise analytical results, let us again consider quasi-special parameters in which $a$ and $b$ are both close to 1, as we studied in Section \ref{Quasi}. We expand all quantities in powers of $\beta$. In order to capture the small $k$ regime, and to keep to a single expansion parameter, we need to express it as a power of $\beta$, namely
\beq
k^2=\beta\,\kappa^2,
\eeq
where $\kappa$ now plays the role of a (re-scaled) wavenumber.

For a generic amplitude $\phi_a$, we take the background solution $\phi(t)$ from Section \ref{pertnohubble} and insert into the equation of motion for the mode functions (\ref{perteom}). Again changing variables from $t$ to $\tau$, using $\tau=\omega\,t$, and working to $\mathcal{O}(\beta)$, leads to
\begin{eqnarray}\label{perter}
\tilde{d}_1(\tau)v_k''(\tau)+\tilde{d}_2(\tau)v_k'(\tau)+\tilde{d}_3(\tau)v_k(\tau)=0,
\end{eqnarray}
\begin{figure}[H]
\centering
\includegraphics[scale=0.22]{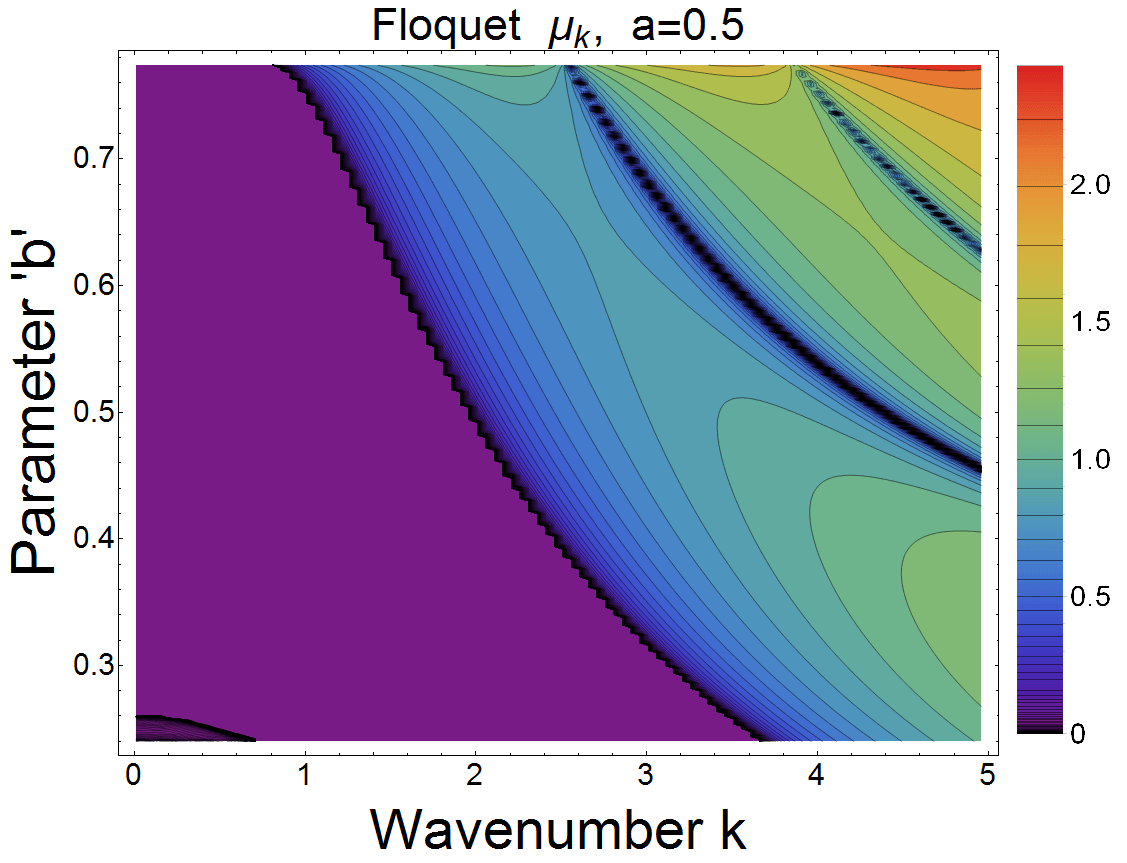}\nolinebreak
\includegraphics[scale=0.22]{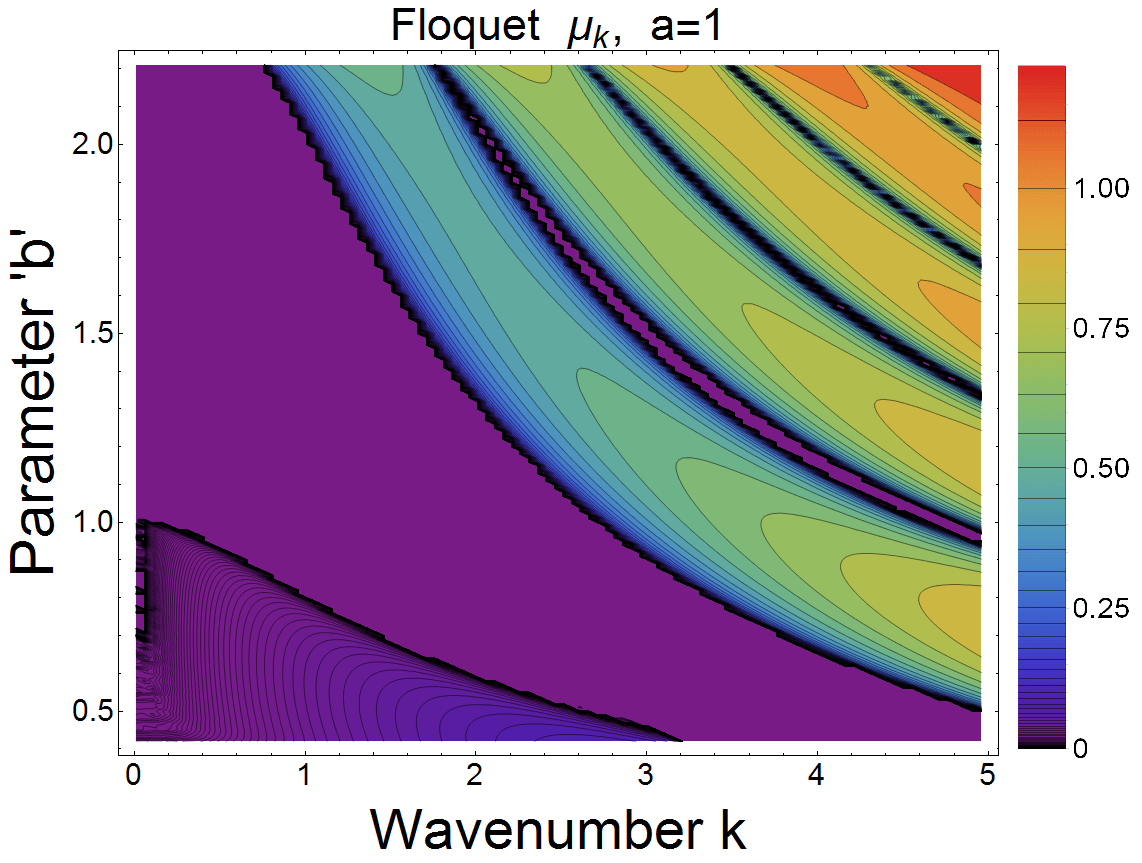}
\includegraphics[scale=0.22]{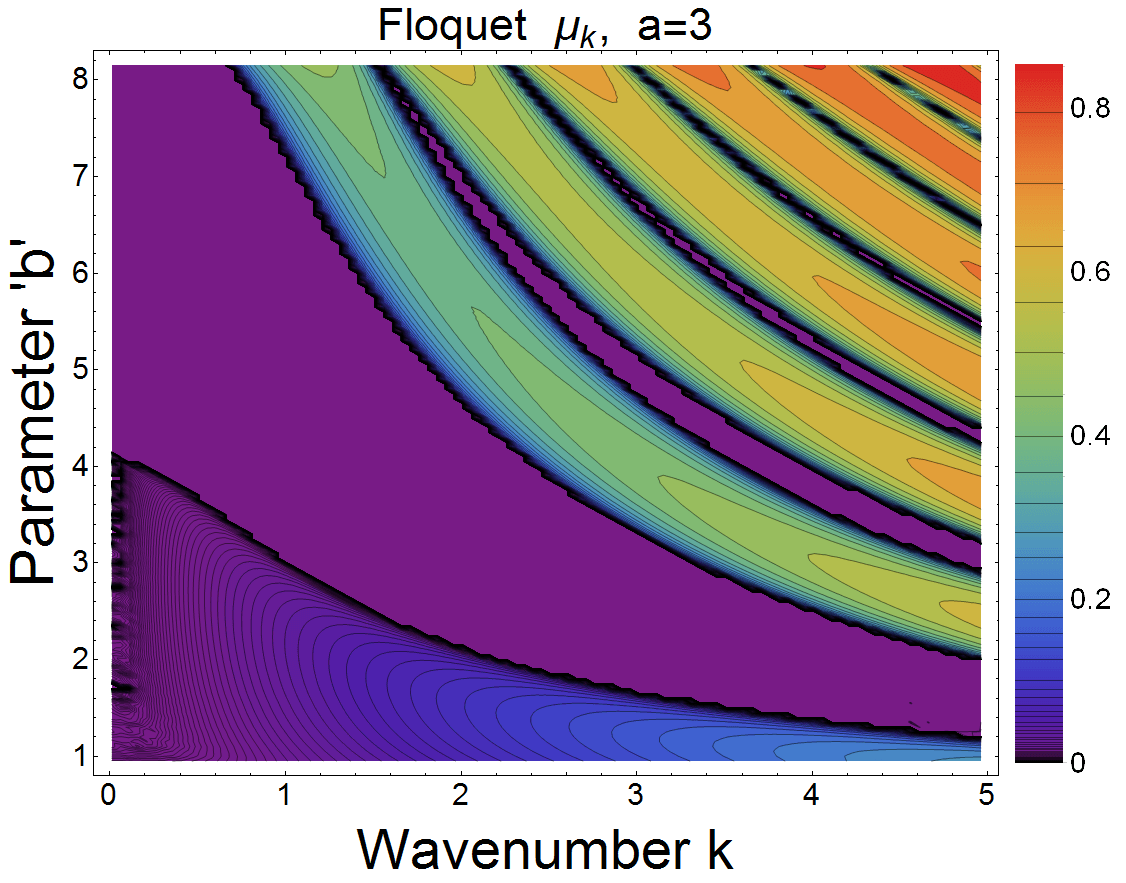}\nolinebreak
\includegraphics[scale=0.22]{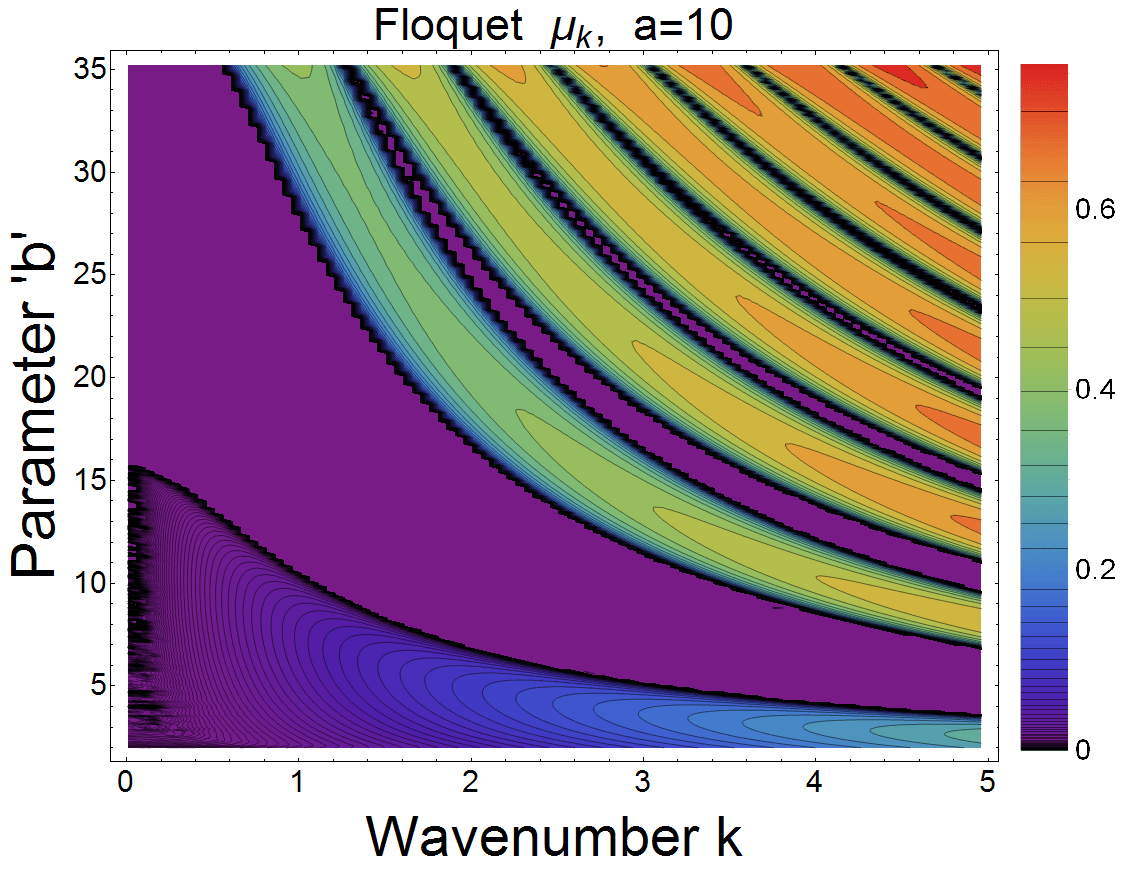}
\caption{Contour plot of the magnitude of real part of Floquet exponent, $|\mathbb{R}(\mu_k)|$, as a function of wavenumber $k$ and parameter $b$, fox fixed $a$. The purple region corresponds to stability with $|\mathbb{R}(\mu_k)|=0$, while the other colors correspond to instability with $|\mathbb{R}(\mu_k)|>0$. Upper left is $a=0.5$, upper right is $a=1$, lower left is $a=3$, lower right is $a=10$.}
\label{contourplots}\end{figure}
\hspace{-0.6cm}where
\begin{eqnarray}
\tilde{d}_i(\tau)=f_i+\tilde{g}_i(\tau),
\end{eqnarray}
with
\beq
\tilde{g}_1(\tau)=g_1\cos(2\tau),\,\,\,\,\tilde{g}_2(\tau)=g_2\sin(2\tau),\,\,\,\,\tilde{g}_3(\tau)=g_3\cos(2\tau),
\eeq
where the prefactors $f_i$ and $g_i$ are given in Appendix \ref{letters}.

It is convenient to substitute the mode functions $v_k(\tau)$ as an expansion in terms of harmonics, weighted by functions $\gamma_n(\tau)$, as follows
\begin{eqnarray}
v_k(\tau)=\sum_{n\in\mathbb{Z}} e^{in\om \tau}\gamma_n(\tau).
\end{eqnarray}
For small wave numbers, the perturbations carry almost the same frequency as the pump and therefore the $\gamma_n(\tau)$ are slowly varying. Hence, to leading order, we can drop second derivatives of $\gamma_n(\tau)$ in their equation of motion. To leading approximation, we can just track the $n=-1$ and $n=1$ terms, which leads to the following $2\times2$ matrix problem
\begin{eqnarray}\label{2x2matrix}
\left( \begin{array}{cc}
\bar{A} & \bar{B}  \\
-\bar{B} & -\bar{A} \end{array} \right)
\left( \begin{array}{c}
\gamma'_{-1}  \\
\gamma'_{1}\end{array} \right)
=
\left( \begin{array}{cc}
\bar{C} & \bar{D}  \\
\bar{D} & \bar{C} \end{array} \right)
\left( \begin{array}{c}
\gamma_{-1}  \\
\gamma_{1}\end{array} \right)
\end{eqnarray}
where
\begin{eqnarray}\label{abcd}
\bar{A}=2i f_1\qquad \bar{B}=-ig_1+\frac{g_2}{2i}\qquad \bar{C}=-f_1+f_3\qquad \bar{D}=-\frac{g_1}{2}-\frac{g_2}{2}+\frac{g_3}{2}.
\end{eqnarray}
By multiplying by the inverse matrix on the left, this becomes
\begin{eqnarray}
\left( \begin{array}{c}
\gamma'_1  \\
\gamma'_{-1}\end{array} \right)
=\bold{G}\left( \begin{array}{c}
\gamma_1  \\
\gamma_{-1}\end{array} \right).
\end{eqnarray}
We find that the eigenvalues of $\bold{G}$ are
\begin{eqnarray}\label{2x2eig}
\lambda=\pm\sqrt{\frac{\bar{C}^2-\bar{D}^2}{\bar{A}^2-\bar{B}^2}}.
\end{eqnarray}
Hence the general solution is of the form
\begin{eqnarray}
\gamma_{\pm}=j_1 e^{\mu_k t}+j_2 e^{-\mu_k t},
\end{eqnarray}
where $\mu_k=\pm\lambda\,\omega$ and $j_1,\,j_2$ are constants that depend on initial conditions. 
Working to the leading non-zero order in $\beta$, we obtain the result
\beq\label{mu2func}
\mu^2_{k}=\frac{(2-3\phi_a^2)\left[3(2-3\aaa)\phi_a^2\kappa^2+(2-11\phi_a^2+15\phi_a^4)\kappa^4\right]}{16(-1+3\phi_a^2)^3}\beta^2.
\eeq
Note
\begin{figure}[t]
\centering
\includegraphics[scale=0.4]{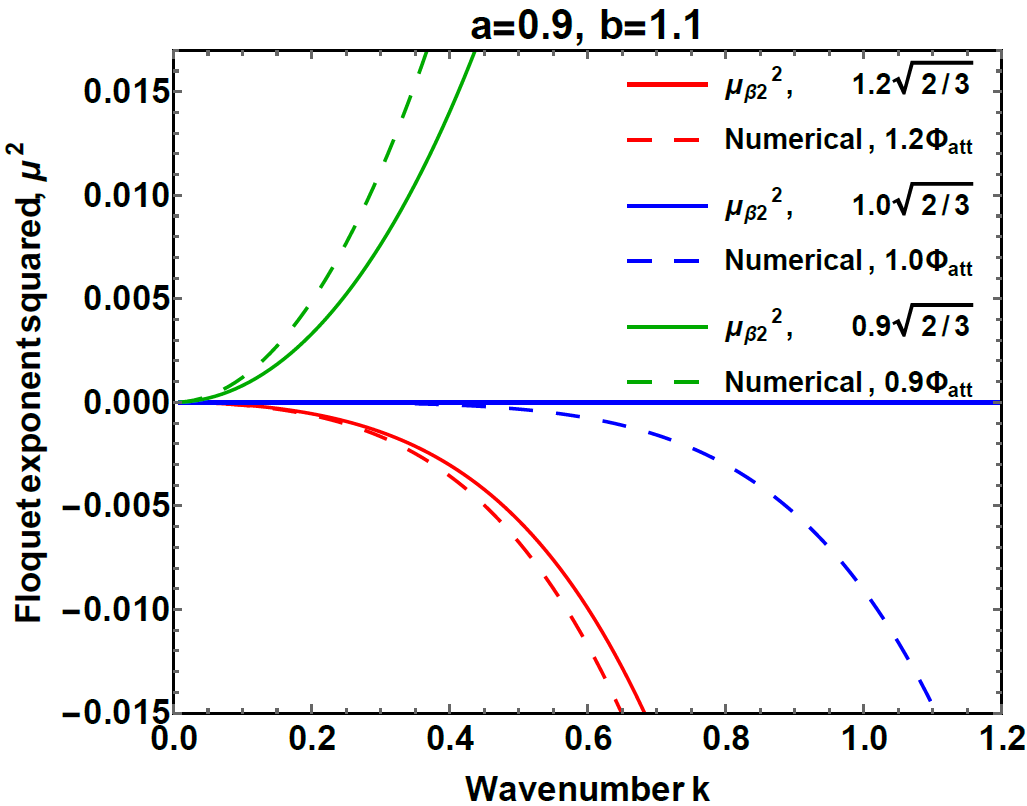}
\caption{Plot of the square of the Floquet exponent $\mu^2_{k}$ as a function of wavenumber $k$, for small $k$, with $a=0.9$ and $b=1.1$.
The dashed curves are the exact numerical results. The solid curves are the approximate analytical result to $\mathcal{O}(\beta^2)$ from eq.~(\ref{mu2func}). The lower (red) curves are for $\phi_a=1.2\,\phi_{a,att}(\Phi_{att})$, the middle (blue) curves are for $\phi_a=\phi_{a,att}(\Phi_{att})$, and the upper (green) curves are for $\phi_a=0.9\,\phi_{a,att}(\Phi_{att})$.}
\label{mu2plot}\end{figure}
that this evidently goes to zero as $k(\kappa)$ goes to zero. Furthermore, we see that at the attractor amplitude (which to leading order is $\phi_a\approx\sqrt{2/3}$), this result vanishes. In Figure \ref{mu2plot} we plot this result for different choice of $\phi_a$, both at the attractor result and away from it.

\subsubsection{Floquet Squared at Attractor}

At the attractor amplitude $\Phi_{att}$ the previous approximation simply gives $\mu_k=0$. Hence we need to go to higher order in powers of $\beta$ in order to obtain the leading non-zero result. 

At next order, we need to track the next harmonic in the analysis. So, using an extension of the previous notation we have
\begin{eqnarray}
d_i(\tau)=f_i+\tilde{g}_i(\tau)+\tilde{h}_i(\tau),
\end{eqnarray}
where 
\beq
\tilde{h}_1(\tau)=h_1\cos(4\tau),\,\,\,\,\tilde{h}_2(\tau)=h_2\sin(4\tau),\,\,\,\,\tilde{h}_3(\tau)=h_3\cos(4\tau).
\eeq
We now need to track $n=-3$, $n=-1$, $n=1$, and $n=3$ terms in the analysis. This leads to the following $4\times4$ matrix problem
\begin{eqnarray}\label{4x4matrix}
{\left( \begin{array}{cccc}
I & J & K & L \\
M & N & O & P \\
-P & -O & -N & -M\\
-L & -K & -J & -I \end{array} \right)
\left( \begin{array}{c}
\gamma'_{-3}(t) \\
\gamma'_{-1}(t) \\
\gamma'_{1}(t)\\
\gamma'_{3}(t) \end{array} \right) 
= \left( \begin{array}{cccc}
A & B & C & D \\
E & F & G & H \\
H & G & F & E\\
D & C & B & A \end{array} \right)
\left( \begin{array}{c}
\gamma_{-3}(t) \\
\gamma_{-1}(t) \\
\gamma_{1}(t)\\
\gamma_{3}(t) \end{array} \right)},
\end{eqnarray}
where the components of the matrices are defined in Appendix \ref{letters}.

\begin{figure}[t]
\centering
\includegraphics[scale=0.5]{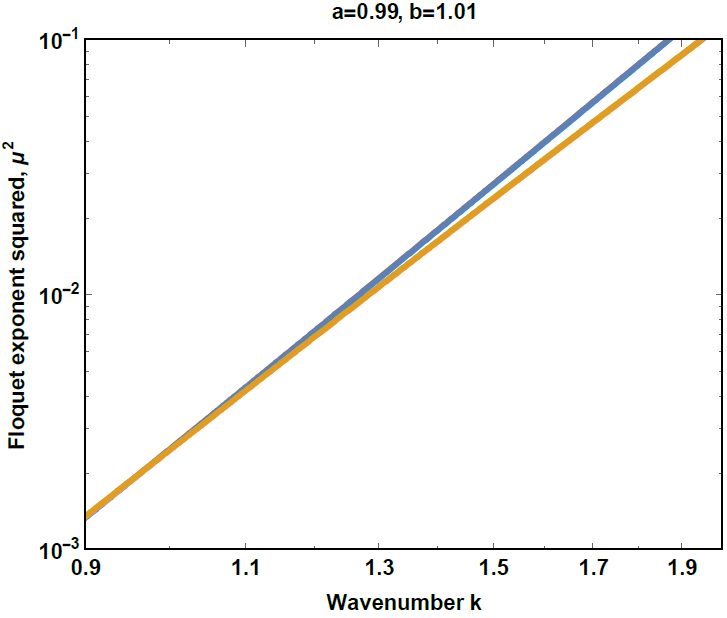}
\caption{Plot of the square of the Floquet exponent $\mu^2_{k}$ as a function of wavenumber $k$, for small $k$, with $a=0.99$ and $b=1.01$ at the attractor amplitude. The lower (yellow) curve is the exact numerical result. The upper (blue) curve is the approximate analytical result to $\mathcal{O}(\beta^3)$ from eq.~(\ref{mu2attr}).}
\label{mu2attplot}\end{figure}

At this higher order, the assumption that all second derivatives of $\gamma_n(\tau)$ can be neglected begins to break down for any amplitude $\phi_a$. However, by focussing on the attractor amplitude, we previously saw that $\mu_k=0$ at the first order and we will see it is small and non-zero at this next order. Since Floquet is small at the attractor, then we can self consistently say that $\gamma_n(\tau)$ is slowly varying. 
By multiplying by the inverse matrix on the left and carefully extracting the eigenvalues to leading non-zero order, we obtain the attractor result
\beq\label{mu2attr}
\mu^2_{k,att}=\frac{1}{864}\left(-6\kappa^4+9\aaa \kappa^4-2\kappa^6\right)\!\beta^3.
\eeq
In Figure \ref{mu2attplot} we compare this approximate analytical result (\ref{mu2attr}) with the exact numerical result, where we see good agreement. 

This analytical result shows that in the low $k$ limit, the criteria for stability ($\mu_k^2<0$) is $\beta>3\,\alpha/2$, i.e., $b>3(a-1)/2+1$. Of course this criteria is only valid if $a$ and $b$ are close to 1, but it proves that there exists a regime of stability for low $k$. The general criteria for stability can be inferred from the earlier Floquet plots in Figure \ref{contourplots}. A quasi-analytical understanding of small $k$ behavior for any parameters can be determined with a type of fluid analysis, to which we now turn.

\subsubsection{Fluid Analysis}\label{fluid}

In the long wavelength limit, the system should re-organize into a new type of effective field theory. Since the background spontaneously breaks time translation symmetry there will be a massless Goldstone mode. In this context, the Goldstone is an adiabatic mode, associated with energy density perturbations, since energy is the conserved quantity associated with time translations. This idea was used recently to examine models of reheating after inflation in Refs.~\cite{Hertzberg:2014iza,Hertzberg:2014jza}.

In an effective theory, the energy density itself can be quantized and separated into a background piece $\rho$ and a fluctuation $\delta\rho$, as follows
\begin{eqnarray}
\hat{\rho}(t,{\bf x})=\rho(t)+\hat{\delta\rho}(t,{\bf x}).
\end{eqnarray}
By using the expression for the Hamiltonian density in eq.~(\ref{ham2}) and expanding to linear order using (\ref{heisenberg}), we find
\begin{eqnarray}
\hat{\delta\rho}(t,{\bf x})=\phi(t)\left(-1+3a\phi(t)^2+3b\dot{\phi}(t)^2\right)\hat{\delta\phi}(t,{\bf x})+\dot{\phi}\left(-1+3b\phi(t)^2+3\dot{\phi}(t)^2\right)\dot{\hat{\delta\phi}}(t,{\bf x}).
\end{eqnarray}

Now in the long wavelength regime, it can be shown that the equation of motion for the energy density perturbations organizes into a type of sound wave equation for a kind of fluid. This is proven in detail in Ref.~\cite{Hertzberg:2014iza} for general Lagrangians. In $k$-space the result is
\begin{eqnarray}
\ddot{\hat{\delta{\rho}}}_k+k^2\, c_s^2\,\hat{\rho}_k=0,
\end{eqnarray}
where $c_s$ is a sound speed, defined in terms of the {\em time-averaged} background pressure $p$ and density $\rho$
\begin{eqnarray}
c_s^2=\frac{\partial\langle p\rangle}{\partial\langle\rho\rangle}.
\end{eqnarray}
The time average is over a period of oscillation of the background. This equation carries a relativistic dispersion relation, as is appropriate for a Goldstone mode.   The corresponding growth is then given by
\begin{eqnarray}
\mu_k=\pm\, i\,c_s\, k.
\end{eqnarray}
So, there is growth for small wavenumbers if and only if the sound speed is imaginary.

\begin{figure}[t]
\centering
\includegraphics[scale=0.4]{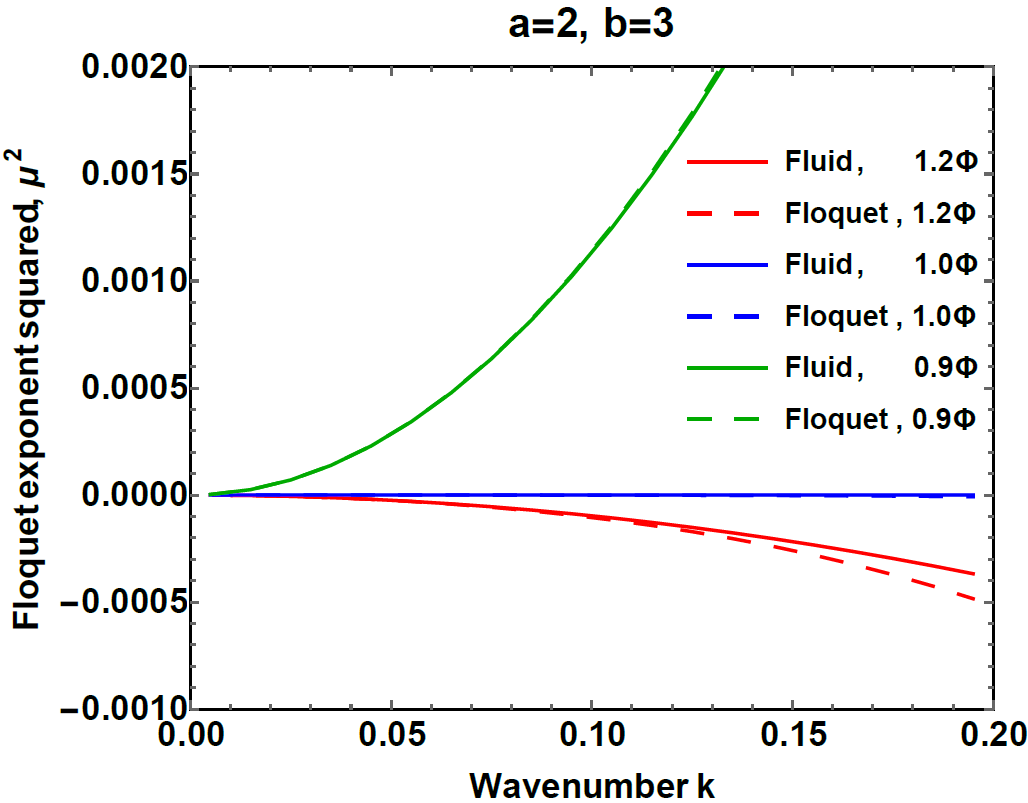}
\caption{The square of the Floquet exponent $\mu_k^2$ as a function of wavenumber $k$, for small wavenumbers, for $a=2$ and $b=3$. The dashed curves are the exact numerical results. The solid curves are the approximate analytical results, using the effective fluid analysis of Section \ref{fluid}. The lower (red) curves are for $\Phi=1.2\,\Phi_{att}$, the middle (blue) curves are for $\Phi=\Phi_{att}$, and the upper (green) curves are for $\Phi=0.9\,\Phi_{att}$.}
\label{csplot}\end{figure}

In Figure \ref{csplot} we plot this quasi-analytical result and compare to the exact numerical result for different amplitudes, focussing on the small $k$ regime. We see very good agreement, including confirmation that the instability vanishes at the attractor to leading order, as we discussed at the end of Section \ref{floquetnohub}.

\section{Cosmological Consequences}\label{Cosmology}

\begin{figure}[t]
\centering
\includegraphics[scale=0.7]{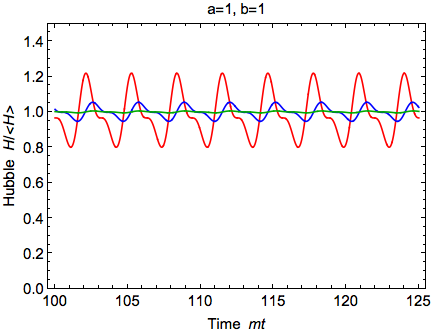}
\caption{Hubble $H$ divided by mean value $\langle H\rangle$ as a function of (dimensionless) time $m t$ for $a=b=1$ and $\Lambda=0$.
The red curve is for $\langle H\rangle/m=1/6$, the blue curve is for $\langle H\rangle/m=1/24$, and the green curve is for $\langle H\rangle/m=1/240$, where $\langle H\rangle=M^2/(6\mpl)$.}
\label{HubblePlot}\end{figure}

For possible applications to cosmology, it is important that at the attractor the effective equation of state is $\langle w\rangle=-1$, from time averaging the pressure and the density over a cycle; see Figure \ref{eqofstate}.
If $H\ll m$ the oscillations in the energy density $\rho$ are very small, as seen in Figures \ref{eqofstate} and \ref{rhotime}.   Therefore it is self-consistent to assume that  $H(t)$ itself has very small oscillations around its mean value. We can see this by self-consistently solving the Friedmann equation. In Figure \ref{HubblePlot} we have numerically solved the system for $H(t)$ for different ratios of $\langle H\rangle/m$.
The behavior of $H(t)$ is found to be of the form
\begin{eqnarray}
H(t)=\langle H\rangle\left(1+f(t)\right),
\end{eqnarray}
where $f$ is a periodic function whose amplitude of oscillation is given by
\begin{eqnarray}
f_{amp}\sim\frac{\langle H\rangle}{m}.
\end{eqnarray}
So the fluctuations in Hubble are 
$\frac{\Delta H}{\langle H\rangle}\sim\frac{\langle H\rangle}{m}$.
When $\langle H\rangle\ll m$, the time derivatives of $H(t)$ will oscillate significantly, but $H(t)$ itself does not. 

Hence we have an approximate de Sitter phase.
In this section we consider some possible cosmological consequences of this behavior. We will reinstate proper units throughout.

For the sake of concreteness and simplicity, let us again look at the special case $a=b=1$ and $\Lambda = 0$. In this case the Lagrangian density can be written in terms of two scales: the mass $m$ and UV cutoff $M$ as follows
\beq
\mathcal{L}=
-{M^4\over 12}
+{m^2\over2}\varphi^2
-\frac{3 m^4}{4M^4}\varphi^4
- \frac{1}{2}(\partial\varphi)^2
+ \frac{1}{4M^4} (\partial\varphi)^4
+\frac{3m^2}{2 M^4}\varphi^2 (\partial\varphi)^2.
\eeq
For the background evolution, we found that the minimum energy configuration has $\rho=0$ ($\Lambda=0$), but this is not the attractor solution. Instead, the attractor solution involves oscillations with positive energy density given by
\beq
\langle\rho\rangle={M^4\over 12}
\eeq
(plus the very small amplitude oscillations around this mean value).
If this field self consistently drives the background Hubble  expansion, then using the Friedmann equation the Hubble parameter becomes
\beq
\langle H\rangle={M^2\over6\mpl}
\eeq
(again plus very small amplitude oscillations around this mean value).
Since the frequency of oscillation is given by
\beq
\omega = m,
\eeq
then in order for oscillations of $\phi$ to be rapid on a Hubble time scale, we need
\beq
{M^2\over6\mpl}\ll m.
\eeq

In this regime $H$ will indeed appear almost constant, and be an approximate de Sitter phase.
In order to drive the present day acceleration, we require $m\gg H_0\approx 10^{-33}$\,eV and $M\approx 3\times 10^{-3}$\,eV.
In order to drive early universe inflation around the GUT scale, we require $m\gg H_0\approx 10^{14}$\,GeV and $M\approx 10^{16}$\,GeV. 

As explained earlier from our study of the Mathieu equation, for $a=b=1$ we do not have any instability for low wave numbers. However, for high wave numbers there is considerable instability. This suggests two simple possibilities for cosmology:
\begin{itemize}
\item There is significant production of $\varphi$ radiation. This may be detectable, depending on how the $\varphi$ particles couple to Standard Model particles. In the centre of the first few instability bands, the growth rate is roughly
\beq
\mu_k\sim m.
\eeq
Furthermore, since the corresponding scale factor will self consistently have some residual oscillations, on top of the overall expansion, this can lead to graviton production. The latter can act as a form of radiation, altering the expansion history of the universe. Once enough quanta are produced, the vacuum may completely decay and we may enter a new phase. This is one possible way to end such a model for inflation. Secondly, as a model of dark energy, it suggests interesting phenomenology for future studies of dark energy. If $m\sim H_0$, such that these resonant effects may only be just beginning to take place, it may lead to observable consequences at the current limit of observation.
\item
One way to avoid the production of $\varphi$ quanta is to imagine that this effective field theory has a particularly low cutoff. If we choose $M\sim m$, then the wave numbers that are being produced, which are $k\sim m$ in the first few instability bands, are beyond the regime of validity of the effective field theory, and therefore we may ignore this behavior. One the one hand, this does not allow for a parametric separation of scales. On the other hand, it is a highly non trivial fact there is no instability for low wave numbers, depending on parameters. This can provide a stable vacuum and would allow for a possibly indefinite oscillatory phase.
\end{itemize}

\section{Conclusion}\label{Conclude}

In this work we have explored a potential new phase of matter in a relativistic scalar field theory. We have found that by allowing for an effective field theory with a negative squared mass term and a negative leading kinetic term, plus higher order stabilising terms in the action, we obtain solutions that oscillate forever in an expanding universe. This complements the ``time crystal" models which focus on motion in the minimum energy state. Here we find that the energy ground state configuration is singular. However, the expanding universe drives the system to a different configuration that oscillates forever in the background, or homogeneous, approximation. Since physical quantities, such as the pressure, oscillate in this ``vacuum" state, we have genuine spontaneous breaking of the time translation symmetry.

Let us note that effective theories of the type described here, which combine friction and non-positive definite kinetic terms to produce time crystal behavior, might also arise in the description of laboratory systems.  

The corresponding effective equation of state of the field is found to be $\langle \rho + p \rangle = 0$ (so $\langle w\rangle=\langle p\rangle/\langle\rho\rangle=-1$). One could imagine that this provides an early time inflationary phase or a late time dark energy phase, depending on parameters.  Although the time average of $p$ is the time average of $-\rho$, we found significant oscillations around this value. This could have interesting consequences.  For a model of late time dark energy, it could be ruled out if the time scale of oscillation were comparable to Hubble. Otherwise, if the time scale of oscillation were very rapid, then it might have escaped observation. 
For a model of early universe inflation, it would be important to compute the spectrum of fluctuations; we leave such explorations for future work.

An important signature of the model is the production of radiation, which can be parametrically excited from the background oscillations. We found that the field can excite high wavenumber modes of $\phi$ quanta. This may destabilse the solution and either lead to a new phase or it could be ruled out, again depending on parameters. It is possible to view the cutoff of the effective field theory to be sufficiently small to ignore this regime of momenta.

More speculatively, one might find these models of spontaneous breaking of time translation symmetry relevant to the creation of the universe and/or the arrow of time problems. Alternatively, it could help inspire new types of model building ideas, for example using non-trivial scalar field dynamics to help address fine-tuning problems, e.g., see \cite{Abbott:1984qf,Graham:2015cka}.

However, the fact that high wave numbers are unstable against parametric excitation suggests that this model does not have a well defined vacuum. This suggests that the model may in fact be pathological and may resist embedding into a UV complete theory that is unitary and causal. We leave these interesting issues for future investigations.

\section*{Acknowledgments}
We would like to thank Babak Haghighat, Daniel Jafferis and Matt Reece for helpful discussions. JSB is supported in part by the Kennedy Scholarship, the Center for the Fundamental Laws of Nature at Harvard University, and GWPI. MH would like to thank support by the Institute of Cosmology at Tufts University and the Center for Theoretical Physics at MIT.  FW is supported by the U.S. Department of Energy under contract No. DE-FG02-05ER41360.

\appendix

\section{Higher Order Results}\label{HigherOrder}

\subsection{Solution at Second Order}\label{Appbeta2}

At $\mathcal{O}(\beta^2)$ the equation of motion is
\bea
\amp \amp\phi_2''+\phi_2\nonumber\\
\amp=\amp\frac{1}{128(-1+3\phi_a^2)^3} \Big{[}\! -128c_2(-1+3\phi_a^2)^2 -9\phi_a^4\left[44-60\aaa+3\aaa^2+48(-2+\aaa (2+\aaa))\phi_a^2\right]\!\cos(\tau)\nonumber\\
\amp\amp \,\,\,\,\,\,\,\,\,\,\,\,\,\,\,\,\,\,\,\,\,\,\,\,\,\,\,\,\,\,\,\,\,\,\,\,\,\,\,\,\,\,\,\,\,\,\,+ 9\phi_a^5\left[36-8\aaa-21\aaa^2+6(-12+\aaa(-4+9\aaa))\phi_a^2\right]\cos(3\tau)\nonumber\\
\amp\amp \,\,\,\,\,\,\,\,\,\,\,\,\,\,\,\,\,\,\,\,\,\,\,\,\,\,\,\,\,\,\,\,\,\,\,\,\,\,\,\,\,\,\,\,\,\,- 27(-2+\aaa)\phi_a^5\left[6-\aaa+10(-2+\aaa)\phi_a^2\right]\cos(5\tau) \Big{]}.
\eea
To avoid secular behavior, the coefficient $c_2$ is
\beq
c_2=-\frac{9\phi_a^4(44-96\phi_a^2+12\aaa(-5+8\phi_a^2)+\aaa^2(3+48\phi_a^2))}{128(-1+3\phi_a^2)^3}.
\eeq
Then, solving for $\phi_2$ we find
\begin{eqnarray}\label{fullphi2}
\phi_2(\tau)\amp=\amp\frac{1}{2048(-1+3\phi_a^2)^3}\left[-18\phi_a^5(36-8\aaa-21\aaa^2+6(-12+\aaa(-4+9\aaa))\phi_a^2)\cos(3\tau)\right.\nonumber\\
\amp\amp\,\,\,\,\,\,\,\,\,\,\,+\left.18(-2+\aaa)\phi_a^5(6-\aaa+10(-2+\aaa)\phi_a^2)\cos(5\tau)+9(-2+\aaa)^2\phi_a^7\cos(7\tau)\right]\!.\,\,\,\,
\label{phi2result}\end{eqnarray}
At the attractor, we can solve for $\phi_a$. We demand that $\langle \pi_\phi\,\dot\phi\rangle=0$ and work to $\mathcal{O}(\beta^2)$ and find
\begin{eqnarray}\label{phiatt}
\phi_2(\tau)_{att}\amp=\amp
\frac{1}{768\sqrt{6}}\left[21(-4+\aaa(4+19\aaa))\cos(\tau)+2(-2+\aaa)(-9(2+3\aaa)\cos(3\tau)\right.\nonumber\\
\amp\amp\,\,\,\,\,\,\,\,\,\,\,\,\,\,\,\,\,\,\,+\left.(-22+17\aaa)\cos(5\tau)+(-2+\aaa)\cos(7\tau))\right].
\end{eqnarray}
Using eq.~(\ref{phi2result}) the relationship between the amplitude $\Phi=\phi_{max}$ and the cosine coefficient $\phi_a$ is
\begin{eqnarray}
\Phi_2&=&\frac{1}{2048(-1+3\phi_a^2)^3}\left[9(-2+\aaa)^2\phi_a^7+18(-2+\aaa)\phi_a^5(6-\aaa+10(-2+\aaa)\phi_a^2)\right.\nonumber\\
&&\hspace{3.8cm}-\left. 18\phi_a^5(36-8\aaa-21\aaa^2+6(-12+\aaa(-4+9\aaa))\phi_a^2)\right]\!.
\end{eqnarray}

\subsection{Matrix Coefficients}\label{letters}

The coefficients to $\mathcal{O}(\beta)$ for arbitrary amplitudes are
\bea\label{oldfs}
f_1&=&(1-3\phi_a^2)-\frac{9(-2\aaa\phi_a^2-2\phi_a^4+9\aaa\phi_a^4)\beta}{8(-1+3\phi_a^2)},\\
g_1&=&\frac{3\phi_a^2(2-6\phi_a^2+3\aaa\phi_a^2)}{4(-1+3\phi_a^2)}\beta,\\
f_2&=&0,\\
g_2&=&-\frac{3\phi_a^2(2-6\phi_a^2+3\aaa\phi_a^2)}{2(-1+3\phi_a^2)}\beta,\\
f_3&=&(1-3\phi_a^2)+\frac{-8\kappa^2-12\phi_a^2+36\aaa\phi_a^2+40\kappa^2\phi_a^2+18\phi_a^4-81\aaa\phi_a^4-48\kappa^2\phi_a^4}{8(-1+3\phi_a^2)}\beta,\\
g_3&=&-\frac{\phi_a^2(18-18\aaa-4\kappa^2-18\phi_a^2+9\aaa\phi_a^2+12\kappa^2\phi_a^2)}{4(-1+3\phi_a^2)}\beta.
\end{eqnarray}
The coefficients to $\mathcal{O}(\beta^2)$ at the attractor amplitude are
\bea
f_1&=&-1+\left(1-\frac{3}{2}\aaa\right)\beta+\frac{1}{64}\left(-76+108\aaa-19\aaa^2\right)\beta^2,\\
g_1&=&(-1+\aaa)\beta+\frac{1}{16}(12-20\aaa+5\aaa^2)\beta^2,\\
h_1&=&\frac{1}{4}(-2+\aaa)\beta+(1-\frac{11}{8}\aaa+\frac{7}{16}\aaa^2)\beta^2,\\
f_2&=&0,\\
g_2&=&(2-2\aaa)\beta+\frac{1}{8}\left(-12+20\aaa-5aa^2\right)\beta^2,\\
h_2&=&(2-\aaa)\beta+\left(-4+\frac{11}{2}\aaa-\frac{7}{4}\aaa^2\right)\beta^2,\\
f_3&=&-1-\frac{1}{3}\kappa^2\beta+\frac{1}{192}\left(-12+60\aaa-27\aaa^2-128\kappa^2+96\aaa\kappa^2\right)\beta^2,\\
g_3&=&\left(-1+2\aaa-\frac{2}{3}\kappa^2\right)\beta+\frac{1}{48}\left(-36+36\aaa-27\aaa^2-40\kappa^2+36\aaa\kappa^2\right)\beta^2,\\
h_3&=&\frac{5}{4}(-2+\aaa)\beta+\left(3-\frac{39}{8}\aaa+\frac{27}{16}\aaa^2\right)\beta^2.
\eea
The corresponding elements of the matrix at the attractor amplitude are
\begin{eqnarray}
&&A=-9f_1+f_3,\,\,\,\,B=\frac{g_3}{2}+\frac{g_2}{2}-\frac{g_1}{2},\,\,\,\,C=\frac{h_3}{2}-\frac{h1}{2}-\frac{h_2}{2},\,\,\,\,D=0,\\
&&E=\frac{g_3}{2}-\frac{9}{2}g_1-\frac{3}{2}g_2,\,\,\,\,F=f_3-f_1,\,\,\,\,G=\frac{g_3}{2}-\frac{g_2}{2}-\frac{g_1}{2},\,\,\,\,H=\frac{h_3}{2}-\frac{3}{2} h_2-\frac{9}{2} h_1,\,\,\,\,\,\,\\
&&I=6i f_1,\,\,\,\,J=i g_1+\frac{g_2}{2i},\,\,\,\,K=\frac{h_2}{2i}-i h_1,\,\,\,\,L=0,\\
&&M=3i g_1-\frac{g_2}{2i},\,\,\,\,N=2i f_1,\,\,\,\,O=\frac{g_2}{2i}-i g_1,\,\,\,\,P=\frac{h_2}{2i}-3i h_1.
\eea

\end{document}